\documentclass[fleqn,1p]{elsarticle}

\usepackage[table]{xcolor}
\usepackage{hyperref}
\graphicspath{ {images/} }
\usepackage{mathtools}
\usepackage{adjustbox}
\usepackage{booktabs}
\usepackage{tikz}
\usepackage{graphicx, subcaption}
\usepackage{multirow}
\usepackage{amsmath}
\usepackage{array}
\usepackage{setspace}
\usepackage{fancyhdr}

\makeatletter
\def\ps@pprintTitle{%
  \let\@oddhead\@empty
  \let\@evenhead\@empty
  \let\@oddfoot\@empty
  \let\@evenfoot\@oddfoot
}
\makeatother

\usetikzlibrary{shapes.geometric, arrows, calc}
\tikzstyle{block} = [rectangle, rounded corners, minimum width=1cm, minimum height=2.25cm,text centered, draw=black]
\tikzstyle{arrow} = [thick,->,>=stealth]

\journal{Digital Signal Processing}

\usepackage{nccmath}
\usepackage{amssymb}

\DeclarePairedDelimiter\floor{\lfloor}{\rfloor}

\biboptions{sort&compress}

\begin{document}

\begin{frontmatter}

\title{Analysis of constant-Q filterbank based representations for speech emotion recognition}


\author[mysecondaryaddress]{Premjeet Singh\corref{corr1}}
\ead{premsingh@iitkgp.ac.in}

\author[mythirdaddress]{Shefali Waldekar}
\ead{swaldeka@gitam.edu}

\address[mysecondaryaddress]{Department of Electronics \& Electrical Communication Engineering \\ Indian Institute of Technology Kharagpur, India }

\author[mymainaddress]{Md Sahidullah}
\ead{md.sahidullah@inria.fr}

\author[mysecondaryaddress]{Goutam Saha}
\ead{gsaha@ece.iitkgp.ac.in}

\address[mymainaddress]{Universit\'{e} de Lorraine, CNRS, Inria, LORIA, F-54000, Nancy, France}

\address[mythirdaddress]{ Department of Electrical, Electronics \& Communication Engineering, GITAM School of Technology, GITAM (Deemed to be University), Bengaluru, India.}

\cortext[corr1]{Corresponding author}

\begin{abstract}
This work analyzes the constant-Q filterbank-based time-frequency representations for speech emotion recognition (SER). Constant-Q filterbank provides non-linear spectro-temporal representation with higher frequency resolution at low frequencies. Our investigation reveals how the increased low-frequency resolution benefits SER. The time-domain comparative analysis between short-term mel-frequency spectral coefficients (MFSCs) and constant-Q filterbank-based features, namely constant-Q transform (CQT) and continuous wavelet transform (CWT), reveals that constant-Q representations provide higher time-invariance at low-frequencies. This provides increased robustness against emotion irrelevant temporal variations in pitch, especially for low-arousal emotions. The corresponding frequency-domain analysis over different emotion classes shows better resolution of pitch harmonics in constant-Q-based time-frequency representations than MFSC. These advantages of constant-Q representations are further consolidated by SER performance in the extensive evaluation of features over four publicly available databases with six advanced deep neural network architectures as the back-end classifiers. Our inferences in this study hint toward the suitability and potentiality of constant-Q features for SER. 
\end{abstract}

\begin{keyword}
Constant-Q filterbank, Constant-Q transform (CQT), Continuous wavelet transform (CWT), Time invariance, Speech emotion recognition (SER).
\end{keyword}

\end{frontmatter}

\thispagestyle{fancy}
\fancyhf{}
\chead{ \footnotesize Accepted for publication in Digital Signal Processing Journal}
\lfoot{ \footnotesize \copyright 2022. This manuscript version is made available under the CC-BY-NC-ND 4.0 license \url{https://creativecommons.org/licenses/by-nc-nd/4.0/}}
\renewcommand{\headrulewidth}{0pt}

\section{Introduction}\label{sec1n}
Speech emotion recognition (SER) is a machine's ability to recognize emotions in a speech sample. With the evolution of \textit{smart} devices, emotion recognition has become an essential facet of artificial intelligence. To master emotion recognition is a long-sought aim of researchers. The ability of machines to automatically gather human sentiment will lead to efficient man-machine interaction. Various applications of SER include assessing a driver's behavior in autonomous driving vehicles, patient monitoring in health-care services, consumer satisfaction in call centers, product analysis~\cite{akccay2020speech,krothapalli2013speech, el2011survey}, etc. Although a considerable amount of progress has been made so far, emotion recognition continues to be one of the most challenging domains in speech signal processing~\cite{picard2000affective, picard2003affective}. This is mainly due to the differences in emotion expression by different individuals. Although few studies suggest that there are no significant differences between how emotions such as \emph{Happy, Angry, Fearful} and \emph{Sad} are expressed by individuals, there are some variations in the intensity of expression among people from different cultures and background~\cite{fischer2004gender, bryant2008vocal, lim2016cultural}.

A general SER system contains an emotion-relevant information (feature) extraction module followed by an emotion-class classifier. In terms of emotion information extraction, the unavailability of a standard feature that promises decent emotion information extraction adds to the challenge around SER. Two types of speech features are prominently used in SER, namely \emph{prosodic} and \emph{spectral} features~\cite{eyben2010towards, gemaps}. Prosodic features mainly include pitch, pitch harmonics, intonation, energy, and speaking rate. Spectral features~\cite{eyben2010towards, gemaps, chen2012speech} include vocal tract resonant frequencies (formants), spectral flux, spectral roll-off, mel-frequency based analysis, etc. Various works which propose specific features for SER generally incorporate techniques to obtain detailed spectral, prosodic, or both spectral-prosodic information~\cite{zhou2009}. The aim here is to extract distinguishing emotional cues. This information is then fed to a classifier that classifies the speech utterances among different emotion classes.

Another popular approach in SER is combining information obtained from both spectral and prosodic speech characteristics~\cite{eyben2010towards}. These methods include extraction of an exhaustive feature set and its statistics over speech segments and whole utterances. They are also termed brute-force methods~\cite{batliner2011whodunnit}. Although these show promising results, \emph{curse of dimensionality} becomes an important issue while handling large number of features. Due to the impressive ability of deep learning methods to find the minima of loss functions, many studies employ \emph{deep neural networks} (DNNs) to self-learn optimum features to either develop an end-to-end SER system~\cite{trigeorgis, huang2017characterizing} or to use the learned features for SER after some post-processing~\cite{mao2014learning, zhang2017speech}.

Deep learning networks provide mathematical models that self-learn an end-to-end solution to a particular pattern recognition problem. This makes deep networks very appealing. However, due to the high complexity, deep networks are onerous to interpret~\cite{lipton2018mythos}, hence making it difficult to obtain an insight into the emotionally relevant characteristics of input speech. Further, due to random initialization of parameters, DNNs converge to different local minima even with identical network architecture and optimization criteria. This raises the question of whether the conclusion about interpretability is consistent for all possible solutions.

Deep learning networks contain many parameters and for their proper training massive amount of data is required~\cite{rolnick2017deep}. As most of the available SER databases are small, transfer learning is frequently used as a substitute approach. However, since the pre-trained models are trained on a completely different database, transfer learning limits the exploitation of the full potential of deep networks. Studies also show that deep networks fail to generalize well in out-of-domain SER scenarios~\cite{parry2019analysis}. Despite all such disadvantages, DNNs perform better in SER by automatically learning features from speech representations as compared to the traditional handcrafted features applied to other machine learning techniques, such as support vector machine (SVM), linear discriminant analysis (LDA), \textit{k}-nearest neighbors (\textit{k}-NN) ~\cite{mao2014learning, zhang2017speech}, etc. 
Therefore, we employ a combination of handcrafted features and deep networks for improved emotion information extraction and enhanced SER performance.

In this work, we use constant-Q filterbank based time-frequency representations with various deep neural network architectures for SER. The time-frequency representation provides the handcrafted descriptor over which the deep neural network further extracts the emotion-relevant information. We analyze the relevance of constant-Q representations from emotion perspective in both time and frequency domains. We then compare constant-Q and mel-scale representations to investigate the effect of different non-linearities on emotion prediction. We also examine and compare the stability of the two features toward time deformations. Additionally, our study shows a striking similarity between the constant-Q feature representations used in this work and first-layer scattering transform coefficients.

In the next section, we review the relevant literature and present our contributions. Section~\ref{tf_rep} elaborates the features used in our experiments. In Section~\ref{comp_tf_rep}, we analyze the differences between time-frequency representation of features in both time and frequency domains. Section~\ref{nn_arch} describes different classifiers employed in this work. The SER databases used in this study are detailed in Section~\ref{databases}. Section~\ref{train_test_eval} outlines the experimentation methods. The results and corresponding discussion are given in Section~\ref{results}, followed by conclusive statements in Section~\ref{sec9}.

\section{Related works and motivation}
\label{sec2}

\subsection{Literature survey}
Before the introduction of deep learning into signal processing, handcrafted features were the default choice of the SER researchers. One of the first seminal works on SER used 17 prosodic features, including speaking rate, voiced region duration, and statistics of pitch~\cite{dellaert1996recognizing}. The work in~\cite{mcgilloway2000approaching} used 32 features post feature selection on various statistics of prosodic features, e.g., mean, standard deviation, skewness, kurtosis, etc. Similarly, for discrimination between stressed and normal speech,~\cite{bou2000comparative} proposed using spectral features with their autocorrelation-based variants and showed significant improvement. Again, considering only spectral features,~\cite{nwe2003speech} used log frequency power coefficients for SER, and they were shown better than mel-frequency cepstral coefficients (MFCCs). In~\cite{bitouk2010class}, statistics of MFCC on three different phoneme classes of speech signal reportedly improved the performance with increased speech segment length. Experiments performed in~\cite{wu2011automatic} showed that modulation spectral features, obtained by applying a separate modulation filterbank on the response of the auditory filterbank, are better in characterising different speech emotions than both MFCC and perceptual linear prediction. In~\cite{wang2015speech}, a discrete Fourier-parameters based model was made for SER. Authors observed that frequency harmonics extracted using Fourier analysis, and their first and second-order derivatives, contain adequate information to discriminate among different emotion classes.

Recently, automatic feature extraction using deep neural networks has gained huge interest because of their ability to learn emotion relevant information from speech signals. Those learned features yield competitive SER performance compared to the traditional handcrafted features. End-to-end SER models were proposed with raw emotional speech and \emph{convolutional neural networks} (CNNs) with \emph{long-short term memory} (LSTM) networks in~\cite{trigeorgis,tzirakis2018end,tang2018end}. Whereas~\cite{zhang2017speech} attempted to learn detailed emotion-related information by providing log mel-spectrogram to the input of CNN and then applying \emph{discriminant temporal pyramid matching}. Similarly,~\cite{ghosh2016representation} used spectrogram of raw speech and glottal waveform as input to stacked denoising autoencoder with bidirectional LSTM (BLSTM) as the classifier. Regarding the suitability of deep learning network architectures for SER,~\cite{fayek2017evaluating} studied 2D CNN, LSTM, and fully-connected (FC) network architectures and reported that 2D CNN network fairs better for SER on IEMOCAP database~\cite{busso2008iemocap}. CNN can be considered a static classifier that jointly processes many speech frames taken at a time. This led to the inference that SER is more dependent on static or utterance-level speech characteristics than dynamic or frame-level information, which is better processed by LSTM networks~\cite{fayek2017evaluating}. Authors in~\cite{parry2019analysis} evaluated the generalization capability of various deep learning architectures in a cross-corpus SER scenario. The study concluded that convolution-based architectures are better for `in the wild' test conditions.

Among different types of handcrafted features used in SER, spectral features are considered to contain a substantial amount of emotionally-rich information. According to~\cite{bitouk2010class}, spectral features ``convey information on both \textit{what} is being said, and \textit{how} it is being said''. A time-frequency representation provides spectral features at different time intervals. In SER, spectrogram, mel-frequency spectral coefficients (MFSCs)\footnote{We refer to mel-frequency spectral coefficients as MFSC in this work. It is also represented as mel-spectrogram in some works. Another equivalent representation is mel-frequency log energy or MFLE.}, and MFCC have been the common choice of time-frequency representation. Spectrogram and its mel-filter variants represent energy distribution across both time and frequency dimensions~\cite{Mika2018}. Such representations provide details about the temporal spread of emotional patterns in the speech signal. However, these time-frequency representations were less effective without post-processing using phoneme class-wise spectral feature extraction in~\cite{bitouk2010class}.

Regarding the frequency domain localization of emotion information in time-frequency representation, several studies have shown that the low-frequency regions of the speech spectrum contain important emotion-related details compared to the high-frequency region~\cite{ williams1972emotions, banse1996acoustic, cowie1996automatic, deb2018multiscale}. These studies reveal the relevance of prosody features for SER and report that the emotions with higher arousal, such as \emph{Angry, Happy,} and \emph{Fear}, have higher average pitch frequency. In contrast, emotions with lower arousal, for example, \emph{Sad}, and \emph{Neutral} have lower pitch values and consistent pitch contours.

Authors in~\cite{bou2000comparative} reported that \emph{Neutral} has better recognition rate around the first formant frequency F1 ($200$-$1000$~Hz) while around the second formant frequency F2 ($1250$-$1270$~Hz), the recognition accuracy of \emph{Anger} is higher. The authors in~\cite{france2000acoustical} found the center frequencies of formants F2 and F3 to reduce in depressed individuals. In~\cite{goudbeek2009emotion}, it was shown that emotions with higher arousal have higher values of mean F1 and lower values of F2, whereas positive valence emotions have higher mean F2. Some discrimination between idle and negative emotions was shown using the temporal patterns of first two formant frequencies in~\cite{bozkurt2011formant}. Authors in~\cite{lech2018amplitude} demonstrated that non-linear frequency scales, such as logarithmic, mel, and equidistant rectangular bandwidth (ERB), positively impact SER performance when compared with linear scale. Hence, the studies performed in SER literature hint at the potential emotion salience of low-frequency speech regions.

\subsection{Motivation and contributions}

\begin{figure*}[t!]
    \begin{subfigure}{0.5\textwidth}
    \centering
        \hbox{\hspace{-1.25cm}\includegraphics[scale=0.45]{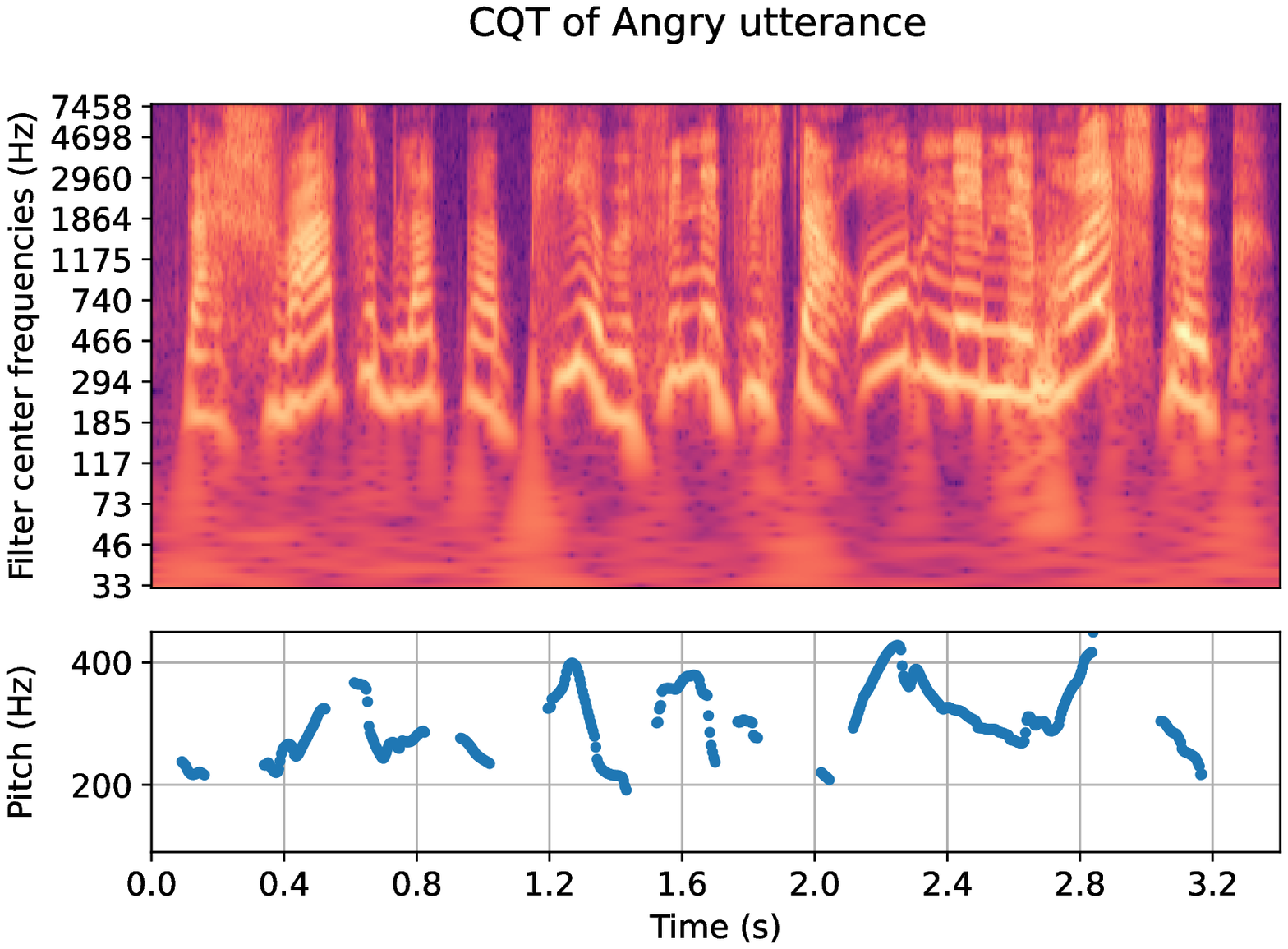}}
    \end{subfigure}
    \begin{subfigure}{0.5\textwidth}
    \centering
        \includegraphics[scale=0.45]{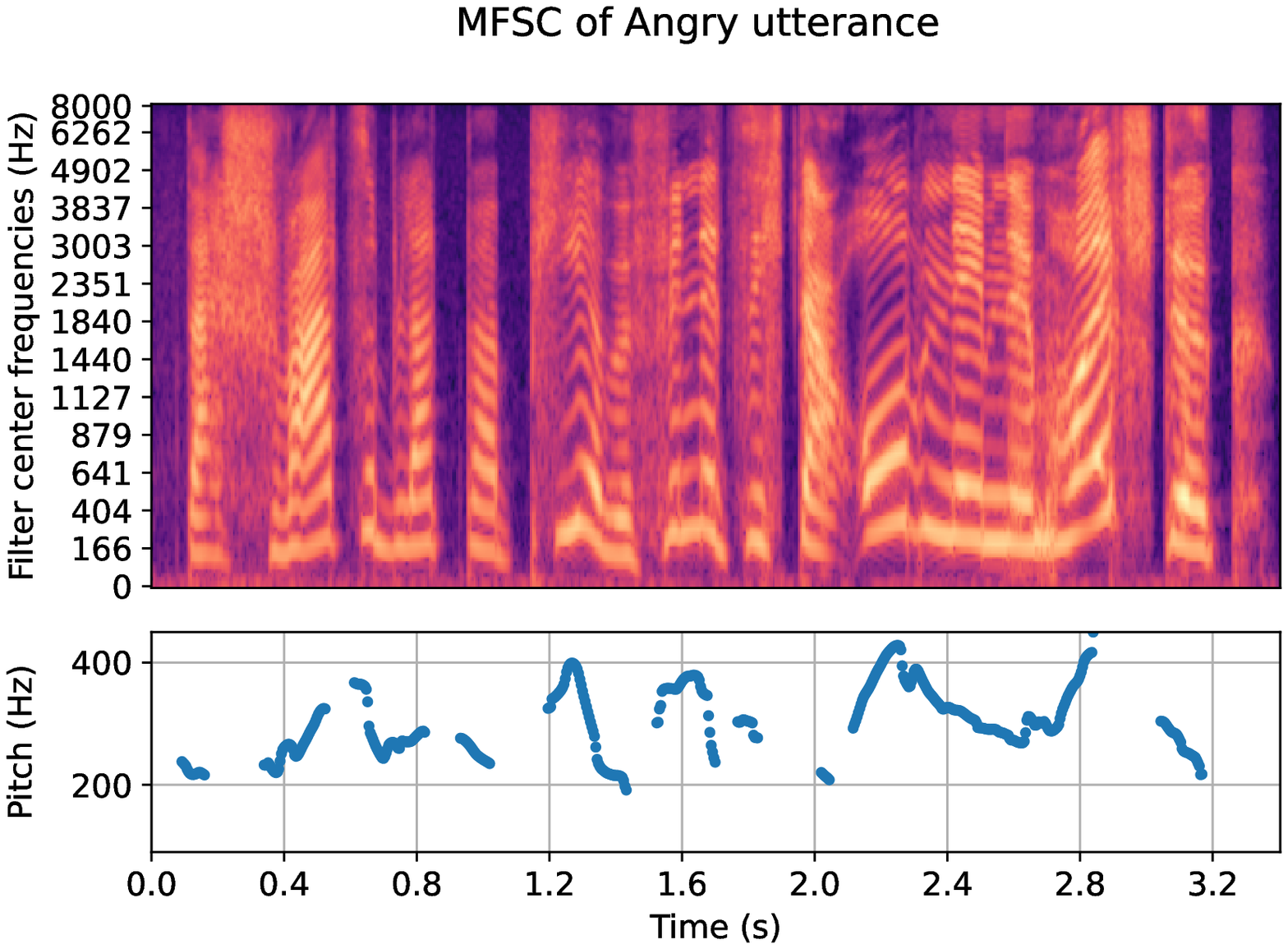}
    \end{subfigure}
    
    \begin{subfigure}{0.5\textwidth}
    \centering
        \hbox{\hspace{-1.25cm}\includegraphics[scale=0.45]{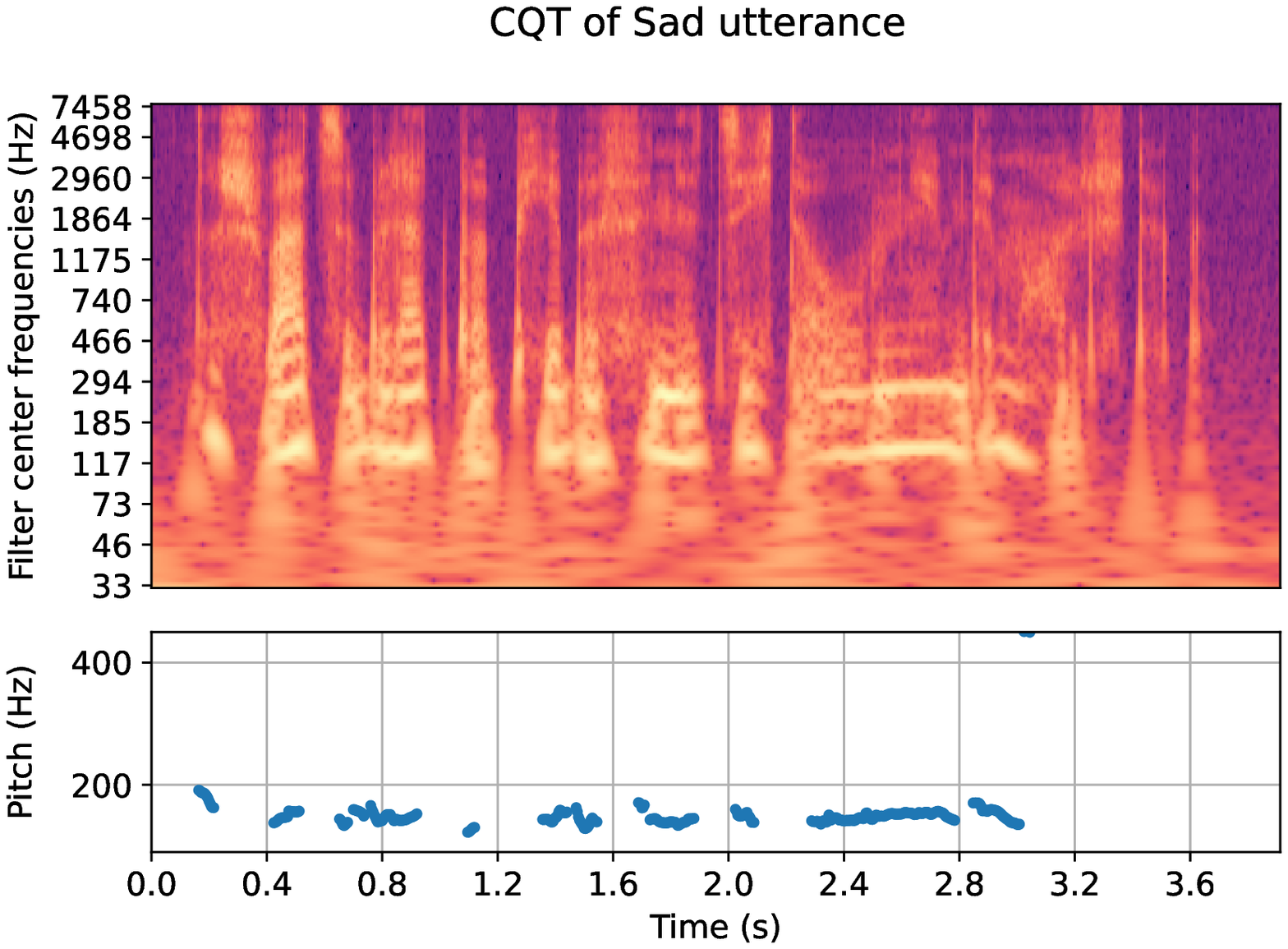}}
    \end{subfigure}
    \begin{subfigure}{0.5\textwidth}
    \centering
        \includegraphics[scale=0.45]{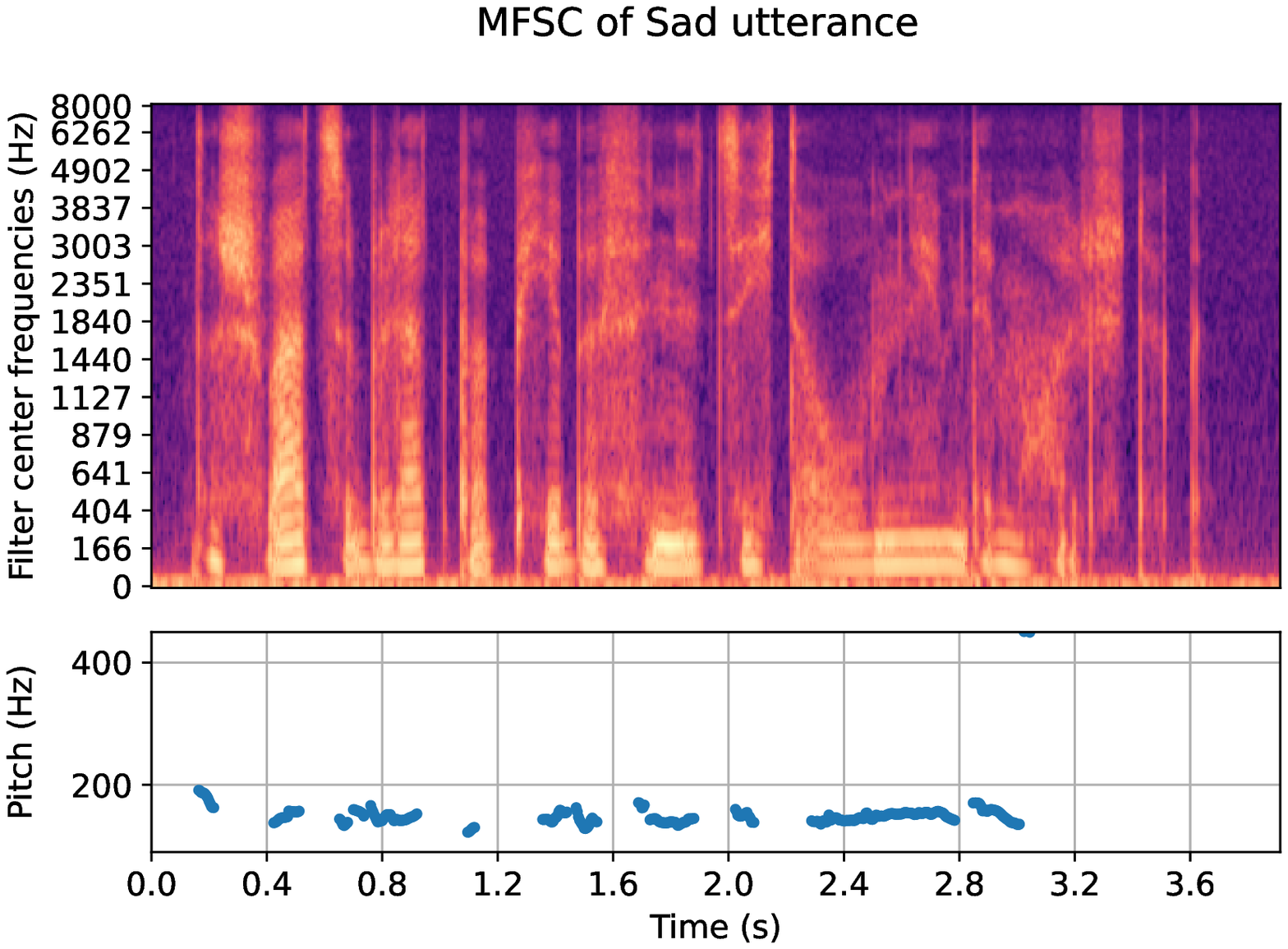}
    \end{subfigure}
    \caption{MFSC and CQT plots of \textit{Angry} and \textit{Sad} utterances of Sentence a05 by Speaker 09 from EmoDB database. CQT representation provides higher number of frequency bins at lower frequencies, hence increasing the low-frequency resolution. This leads to improved pitch resolution in CQT as compared to MFSC, especially in \emph{Sad} emotion where pitch values are low. Another observation in CQT is the smearing of low-frequency components in time, increasing time-invariance.}
    \label{fig:Specs}
\end{figure*}

The existing SER literature draws attention to the importance of low frequencies of speech signals for emotion recognition. To exploit this characteristic, a non-linear time-frequency representation is required that can emphasize the low-frequency regions. The mel-scale warping in well-known mel-frequency based analysis introduces a logarithm-based non-linearity which provides some emphasis on lower frequencies. To further improve the resolution, we propose the use of constant-Q filterbank based non-linear scale for SER. The constant-Q transform (CQT) applies constant-Q filters to offer higher frequency resolution in low-frequency regions and higher time resolution in high-frequency regions. We hypothesize that, since the pitch harmonics and lower formants, which play a major role in emotion discrimination, reside in the speech spectrum's low-frequency regions, having a higher resolution in this region would capture emotion-related information more efficiently. Authors in~\cite{chandra} also argue that in CQT, the pitch information is visible and harmonics are well separated. Figure~\ref{fig:Specs} compares CQT and MFSC time-frequency representations for two different emotions from EmoDB database. CQT was originally proposed for music processing~\cite{brown1991calculation}, after which it was successfully applied to other audio processing applications, like anti-spoofing~\cite{todisco2017constant, pal2018synthetic}, speaker verification,~\cite{delgado2016further} and acoustic scene classification~\cite{lidy2016cqt, waldekar2018classification}. In~\cite{tang2018end}, CQT was also studied for SER, but no significant improvement was observed. The reason could be the end-to-end model's inability to exploit CQT completely or the inappropriate choice of CQT computation parameters in the experiments.

Another transform that provides constant-Q filter based structure is the continuous wavelet transform (CWT)~\cite{rioul1991wavelets}. CWT also gives varying frequency resolution, similar to CQT, by utilising different scale values of the wavelet basis function. Hence, we also use the CWT time-frequency representation and compare it with MFSC representation for SER. The difference in CQT and CWT then lies only in how time-frequency representation is computed. This difference also helps us in analysing the importance of time-invariance in feature representation for SER. Although CQT is very similar to CWT, the former provides non-redundant features and is, therefore, better suited for a varying resolution time-frequency representation. Also, the CWT has been found superior than mel-filter based techniques for SER in different works, e.g.,~\cite{huang2015, Ntalampiras2012, wang2015time}, and~\cite{shegokar}. 

In this work, we extend our preliminary study performed in~\cite{singh2021non} with an in-depth analysis focused on the advantages of constant-Q representations for SER and its comparison with the mel-scale features. We also extensively evaluate features on different databases, using six different DNN architectures. The following are the main contributions of this paper:

\begin{itemize}

     \item Detailed time-domain analysis of CQT time-frequency representation for SER and comparison with MFSC.

     \item Time-domain analysis of CWT for SER and its comparison with CQT and MFSC.

     \item Frequency-domain investigation of differences between mel and constant-Q filterbank representation.

     \item Comparison of the performances obtained with different deep neural network architectures for SER.
     
\end{itemize}

\section{Description of time-frequency representations}
\label{tf_rep}

\subsection{Constant-Q transform}

\begin{figure*}[t!]
    \centering
    \hbox{\hspace{-1cm}\includegraphics[scale=0.4]{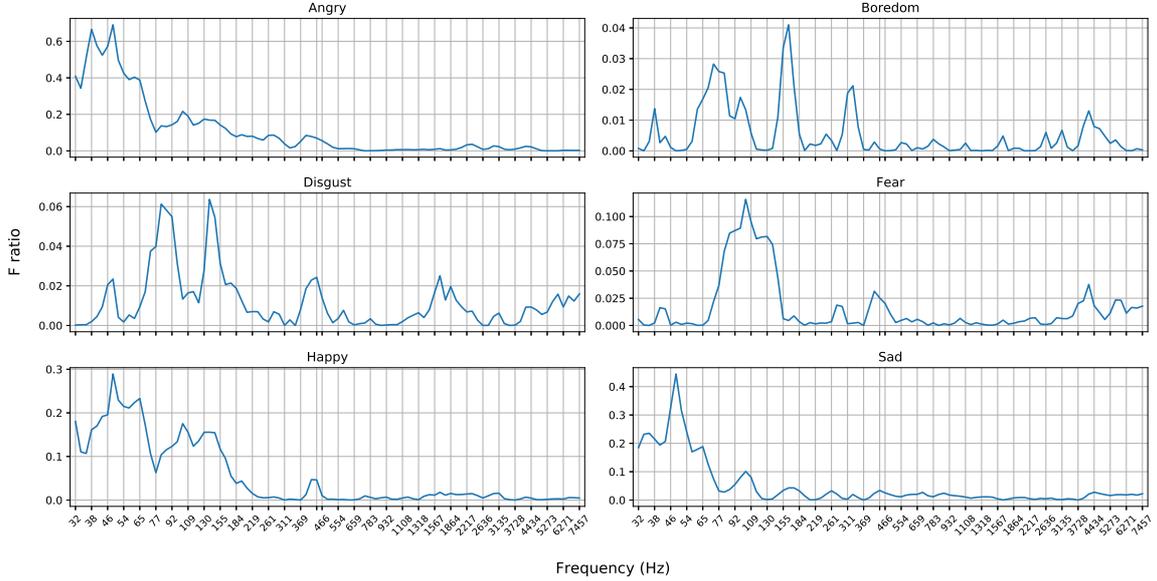}}
    \caption{CQT F-ratio for different frequency bins over EmoDB database.}
    \label{CQT_F_ratio}
\end{figure*}

The constant-Q transform of a time-domain signal $x[n]$ is given as~\cite{schorkhuber2010constant},

\begin{equation}
    X^{CQT}[k,n]~=\sum_{j=n-\floor{N_k/2}}^{n+\floor{N_k/2}}~x(j)a_k^*(j-n+N_k/2)
    \label{eq:CQT}
\end{equation}

\noindent where, $k$ denotes the CQT frequency index, $\floor{.}$ denotes the rounding-off to nearest integer towards negative infinity and $a_k^*(n)$ is the complex conjugate of the CQT basis function for $k$\textsuperscript{th} CQT bin. The CQT basis, or the time-frequency \emph{atom}, is a complex time-domain waveform given as,

\begin{equation}
    a_k(n)~=~ \frac{1}{N_k}w\left(\frac{n}{N_k}\right) exp\left[-i2\pi n\frac{f_k}{f_s}\right]
\end{equation}

\noindent where, $f_k$ is the center frequency of $a_k$, $f_s$ is the sampling frequency, and $w(n)$ is the window function. In this work, \emph{Hann} window is used for CQT calculation. The non-linear placement of $f_k$ in CQT is given by the relation $f_k~=~f_{min}2^\frac{k-1}{B}$ with $B$ being the number of frequency bins used per octave of frequency and $f_{min}$ is the frequency of the lowest bin. This bin placement is inspired by the equal-temperament scale used in music analysis~\cite{todisco2017constant}. The constraints of equal-temperament scale and constant-Q factor of filters result in window length varying over frequency index $k$, given by,

\begin{equation}
    N_k~=~\frac{f_s}{f_k}Q.
\end{equation}

\begin{figure*}[t!]
    \centering
    \hbox{\hspace{-1cm}\includegraphics[scale=0.4]{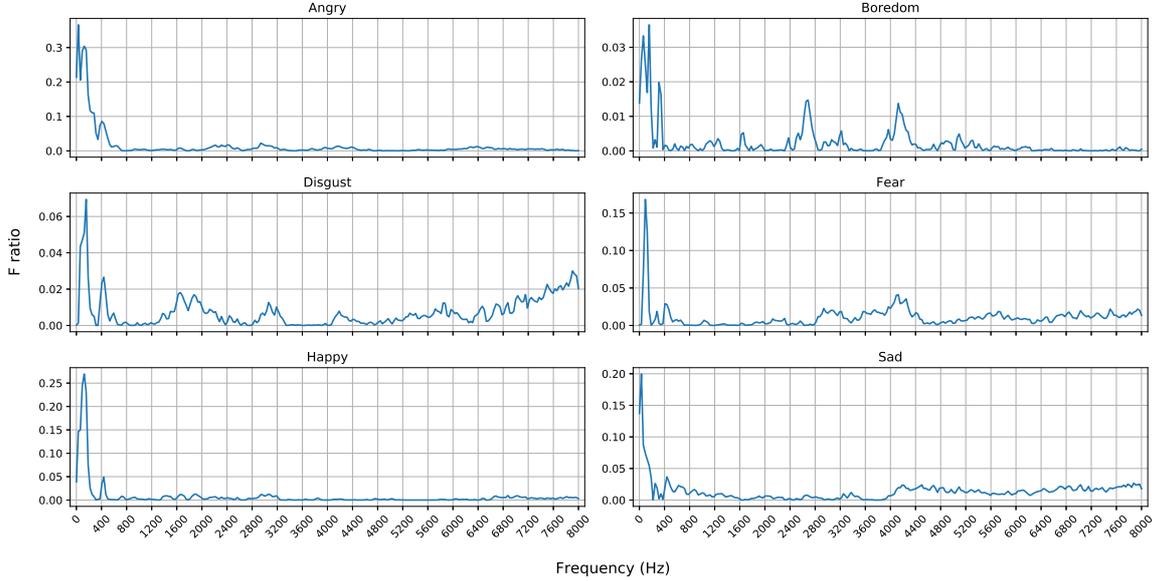}}
    \caption{Spectrogram F-ratio for different frequency bins over EmoDB database.}
    \label{Spec_F_ratio}
\end{figure*}

Equation~\ref{eq:CQT} describes the CQT computation as convolution of the \emph{atom} with every input signal sample. However, to reduce redundancy in feature and computational complexity,~\cite{schorkhuber2010constant} proposed a CQT implementation which uses hop-length as a parameter to shift the constant-Q atom (or constant-Q basis) by a specific number of samples and hence provides time-frames in the time-frequency representation. Further improvements in CQT computation mentioned in~\cite{schorkhuber2010constant} are,

\begin{itemize}
    \item Computation of frequency-domain inner product (correlation) between the signal frame and time-frequency atom (spectral kernel) instead of time-domain computation.
    \item Use of only non-zero values of sparse spectral kernels for computation.
    \item Octave-wise transformation, starting from the highest octave followed by downsampling and low-pass filtering of the signal to obtain lower octaves.
\end{itemize}

\noindent These computation steps differentiate CQT from CWT and provide a non-linear time-frequency representation with reduced redundancy and computational complexity. In our experiments, we used the CQT implementation provided in the \emph{LibROSA}\footnote{\url{https://librosa.github.io/}} toolbox which uses the above mentioned computational improvements.

\begin{figure*}[t!]
    \centering
    \hbox{\hspace{-1cm}\includegraphics[scale=0.4]{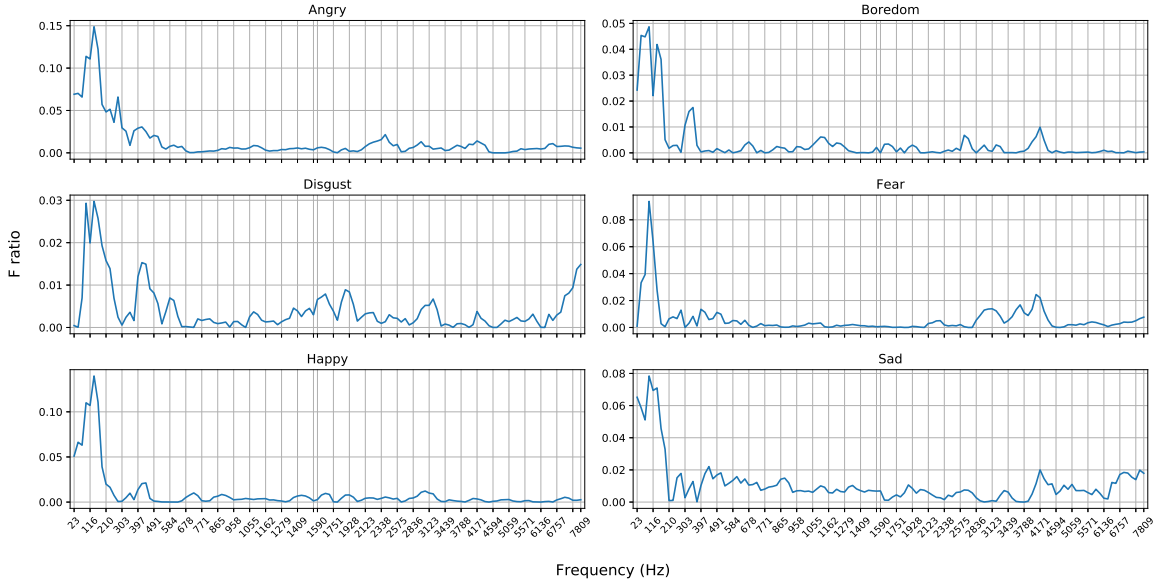}}
    \caption{MFSC F-ratio for different frequency bins over EmoDB database.}
    \label{MFSC_F_ratio}
\end{figure*}

When discrete cosine transform (DCT) is applied on CQT values after uniform resampling, the obtained coefficients are called constant-Q cepstral coefficients (CQCC)~\cite{todisco2017constant}. However, when DCT is applied without resampling of CQT coefficients, we obtain constant-Q coefficients (CQC)~\cite{yang2020improving}. To get an estimate of class-separability of time-frequency representations, we performed the F-ratio analysis (as given in~\cite{nicholson1997evaluating}) on frequency bins obtained from CQT and MFSC. Figure~\ref{CQT_F_ratio} and~\ref{MFSC_F_ratio} show the F-ratio statistic of CQT and MFSC obtained at different frequency bins for different emotions with respect to \textit{Neutral} emotion on EmoDB database. The higher F-ratio values at low-frequency bins of CQT and MFSC show the presence of more emotion discriminative information at these bins. The figures also indicate that CQT-spectrogram has higher percentage of discriminative coefficients on an average due to higher resolution in low-frequency regions. Similarly, Fig.~\ref{Spec_F_ratio} shows the F-ratio plot for STFT (short-time Fourier transform) based spectrogram. Although some emotion classes in spectrogram has higher F-ratio values, the percentage of such discriminative coefficients is low in total. However, most of the higher F-ratio coefficients are again observed at low frequencies.

CQT was first introduced for music analysis owing to its ability to model an equal-tempered frequency scale followed in Western music~\cite{todisco2017constant}. Studies performed in psychology also explain some relationship between music training and emotion perception. Individuals with music expertise can better perceive emotions from speech~\cite{lima2011speaking, good2017benefits}. As CQT is a well-matched representation for music, this could also explain its suitability for SER. Although the CQT configuration found best for SER in our previous contribution (\cite{singh2021non}) differs slightly from that used in music analysis, the division of complete frequency range in octaves with equal number of bins supports the similarity between music perception and SER.

\begin{figure*}[t!]
    \centering
    \hbox{\hspace{-1cm}\includegraphics[scale=0.4]{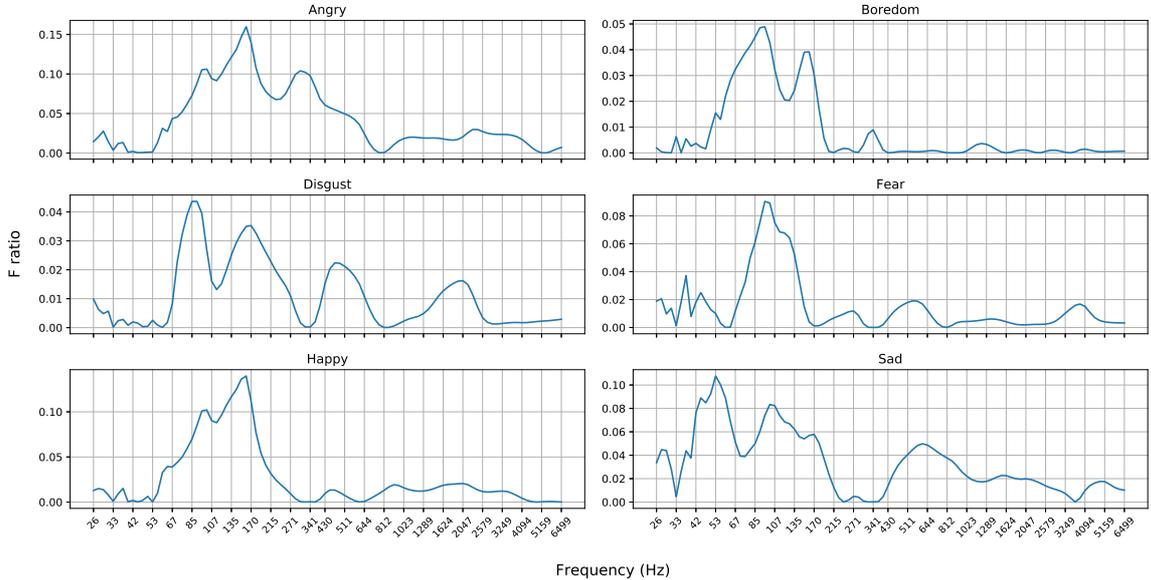}}
    \caption{CWT F-ratio for different frequency bins over EmoDB database.}
    \label{Wav_F_ratio}
\end{figure*}

\subsection{Continuous wavelet transform}
CWT is another transform that provides time-frequency representation of the input signal with varying frequency resolution~\cite{vetterliwavelet}. CWT of a signal $x(t)$ is given as,

\begin{equation}
    \psi_{a,b}(t) = \frac{1}{\sqrt{|a|}} \int_{-\infty}^{\infty} f(t)  \psi  \left(\dfrac{t-b}{a}\right) dt
    \label{eq:CWT}
\end{equation}

\noindent where, $\psi(t)$ is the wavelet basis function while $a$ and $b$ refer to the scale and translation values of the basis, respectively. The large scale values of basis are used to extract slowly varying attributes, whereas smaller scale values can capture minute details of the signal $x(t)$. The collection of wavelet basis with selected scale values also provides a constant-Q factor filterbank representation~\cite{vetterliwavelet}. Moreover, the scale values in CWT can also be selected to obtain an equal-temperament representation similar to CQT. Thus, CWT also provides high frequency resolution at low frequencies and high time resolution at high frequencies. As described by Eq.~\ref{eq:CWT}, CWT coefficients are computed for every integer value of dilation $u$ leading to a highly redundant and computationally extensive coefficient representation for different scale values. This makes the use of CWT as time-frequency representation less efficient. As the CNN requires the input in image (2-dimension) form, we performed sliding window fashioned summation of CWT coefficients to reduce the size of image (time-frequency representation) in time axis, making it more suitable to be used with CNN. However, the issue of redundancy and computation time remains unaffected. Our coefficient summation approach is also based on windowing performed in MFSC to obtain frames in the corresponding time-frequency representation.

In our CWT implementation, we used \emph{complex Morlet} wavelet pertaining to its relatedness with human perception of vision and sound~\cite{shegokar}. We used dyadic scale values ($2^a$, where $a \in \mathbb{R}$) in CWT for fair comparison with CQT time-frequency representation. The discrete wavelet transform (DWT) provides orthogonal wavelet bases which provide uncorrelated time-frequency features at their disposal. Also, since CNNs are efficient in extracting correlation across time and frequency, redundancy due to correlation among features available in CWT should assist in better capturing of correlation across CWT coefficients for SER. Hence, we exclude DWT from our experiments. Figure~\ref{Wav_F_ratio} shows the F-ratio plot of frequency bins of CWT for different emotion classes in EmoDB database. The plot shows that CWT also has frequency bins which are discriminative (have greater F-ratio values). However, the magnitude of F-ratio values are less for every emotion indicating lesser discriminative characteristics of CWT coefficients.










\subsection{Mel-frequency spectral coefficients}
\label{sec_SERfeatures}

Mel-frequency based features, specifically MFSC and MFCC, are widely used in speech signal processing. These features are computed by framing and windowing the speech into short segments followed by computation of power spectrum of every frame~\cite{SAHIDULLAH2012}. This provides a time-frequency representation with number of time frames on one dimension and number of frequency bins on the other. A mel-frequency based filterbank is then applied on the frequency response of every frame to obtain an output coefficient from each filter and hence introduce perception based non-linearity on the time-frequency representation. 

Although the non-linearity in mel-scale is based on human sound perception, it was mainly designed to recognize phonemes in speech. Also, mel filterbank provides poor low-frequency resolution as compared to constant-Q filterbank. As a result, the time-frequency representation generated using mel-scale provides less emphasis on the emotion salient frequency regions of speech signal. The lower F-ratio values of MFSC in Fig.~\ref{MFSC_F_ratio} throughout the complete frequency range further indicates the inferiority of MFSC to discriminate among different emotions classes.

We also performed the F-ratio analysis of CQT, CWT, MFSC, and spectrogram feature on IEMOCAP and eNTERFACE. We found higher emotion discriminability of low-frequencies with these databases as well. However, here we report F-ratio analysis only over EmoDB to prevent text redundancy and to have F-ratio plots made with audio files containing the same contextual information (i.e., transcript).

\section{Comparison between time-frequency representations}
\label{comp_tf_rep}

To understand the effect of filters applied on speech with non-linearly spaced center frequencies from SER perspective, we performed both time and frequency domain comparison between mel-scale and constant-Q based time-frequency representations. In time-domain analysis, we analyzed the time-warp stability and time-shift invariance of MFSC by reproducing the analysis performed in~\cite{anden2014} and then extended it for analysis on constant-Q filterbank based features.

\subsection{Time-domain representation of MFSC}

Consider a signal $x(t)$, its time-shifted version is given by $x(t-c)$. The relation between Fourier transform of the original and the shifted version of the signal is,

\begin{equation}
    X_c(\omega)~=~e^{-j\omega c}X(\omega).
\end{equation}

\noindent Applying modulus on both sides, we obtain,

\begin{equation}
    |X_c(\omega)|~=~|X(\omega)|.
    \label{ft_inv}
\end{equation}

\noindent Equation~\ref{ft_inv} shows that the modulus of the Fourier transform remains stable to time shift $c$ under the condition that $c~<<~T$, where $T$ is the duration of the window over which the Fourier transform is computed. Hence, Fourier transform is inherently stable but only for duration $T$. 

Consider a time-warped (deformed) version of the signal given by $x_\tau(t)~=~x(t-\epsilon t)$, with $0 < \epsilon \ll 1$. The modulus of Fourier transform of the time warped signal is given as,

\begin{equation}
    |X_\tau (\omega)| = \left|\frac{1}{1-\epsilon}X\left(\frac{\omega}{1-\epsilon}\right)\right|.
\end{equation}

\noindent The factor $\frac{1}{1-\epsilon}$ in the frequency term leads to a shift of the frequency components at $\omega'$ by a factor $\epsilon| \omega'|$. This effect gets more prominent at higher values of $\omega$. Hence, the Fourier transform is stable to time-shifts but not to time-warping. Such effect would cause the same emotion utterance of different speakers (e.g., same utterance of a male and a female speaker) to have different frequency attributes making the classification process more difficult. Also, in STFT, the utterance is divided into frames of smaller duration (for example, $20$~ms) and then the modulus of Fourier transform is computed for every frame. The STFT, therefore, remains invariant to time-shifts existing only under $20$~ms duration. However, since speech segments with length greater than $250$~ms contain the information required for emotion prediction~\cite{zhang2017speech}, the STFT fails to capture it efficiently. Therefore, the STFT (spectrogram) time-frequency representation is not an effectual representation from SER perspective.

In mel-scale based representations, a mel-filterbank is applied on the computed STFT to obtain MFSC/MFCC. The mel-filterbank contains filters with logarithmically spaced center frequencies. Let the signal frame centered at time $t$ be given as $x_t(u)~=~x(u)\phi(u-t)$, where $u$ is the time index and $\phi$ is the framing window. In MFSC, the Fourier transform of the signal frame ($X_t(\omega)$) is multiplied with Fourier transform of every mel-filter ($\psi_\lambda(\omega)$) and then summed to obtain a single coefficient at the output of every filter. Mathematically, this is given as,

\begin{equation}
Mx(t,\lambda) = \frac{1}{2\pi} \int |X_t(\omega)|^2|\psi_\lambda (\omega)|^2 d \omega
\label{mfsc_freq}
\end{equation}

\noindent where, $\lambda$ is the support or center frequency of the mel-filter, $t$ is the center of the time frame and $Mx(t,\lambda)$ is the corresponding MFSC . Equation~\ref{mfsc_freq} can also be considered as averaging in frequency domain. Converting the multiplication in frequency-domain in Eq.~\ref{mfsc_freq} into convolution in time-domain, we get,

\begin{align*}
\begin{split}
    Mx(t,\lambda) & = \int |x_t * \psi_\lambda(v)|^2dv
    \\
    & = \int \left|\int x(u)\phi(u-t)\psi_\lambda(v-u)du\right|^2 dv.
\end{split}
\end{align*}

Since the filter support $\lambda$ of mel-filters in time is generally smaller than the support of $\phi$, we have, 

\begin{align}
\begin{split}
    Mx(t,\lambda) & \approx \int \left| \int  x(u)\psi_\lambda (v-u)du \right|^2 |\phi(v-t)|^2dv
    \\
    & = |x*\psi_\lambda|^2*|\phi|^2(t).
\end{split}
\label{mfsc_final}
\end{align}

Equation~\ref{mfsc_final} shows that averaging performed in frequency-domain (Eq.~\ref{mfsc_freq}) is equivalent to time-domain averaging by filter $\phi$. Also, the averaging in frequency domain in Eq.~\ref{mfsc_freq} provides one single coefficient representation of a band of frequencies (defined by bandwidth of mel-filters). This averaging makes the MFSC representation less susceptible to frequency shifts and hence stable to time-warps~\cite{anden2014}. Thus the MFSC representation is invariant to both deformation and time-shift but only for a specific frame duration (here, $20$~ms). Therefore, although stable, MFSC is still incapable of capturing long time scale information, which is essential for SER~\cite{zhang2017speech}. The left most column of Fig.~\ref{block_diagram} shows a pictorial representation of different steps involved in MFSC computation.

\subsection{Time domain representation of CQT}
\label{time_domain_ref}

\begin{figure}[t!]
    \centering
    \includegraphics[scale=0.7]{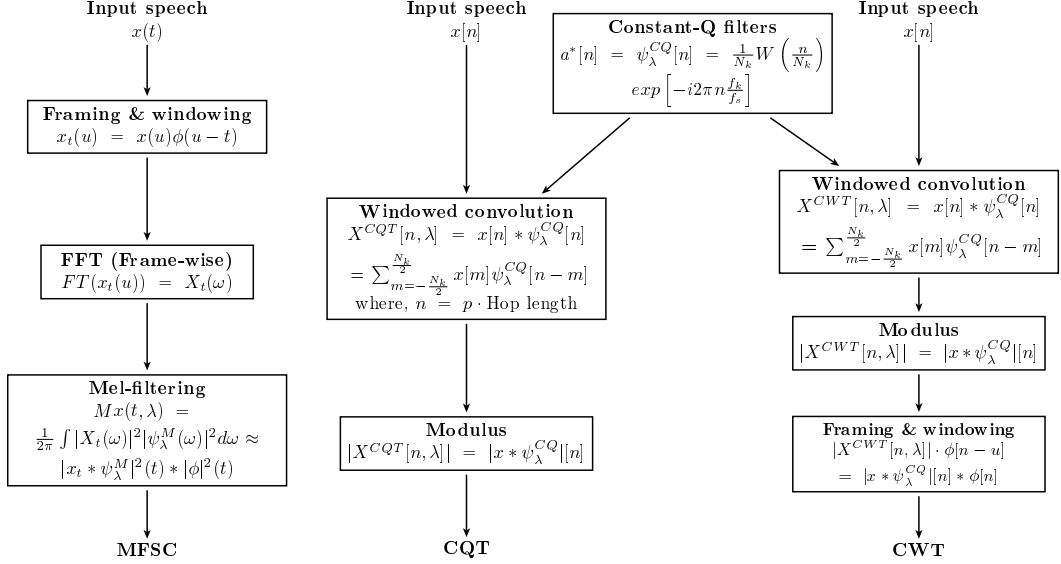}
    \caption{Block diagrams showing time-domain formulation of different features used in this work. Here, $\psi_{\lambda}^{M}$ and $\psi_\lambda^{CQ}$ represent mel and constant-Q filters with frequency support $\lambda$, $\phi(t)$ is the averaging window, $p \in \mathbb{Z}^{+}$,~$N_k$ and $f_k$ are the time-span and frequency of the $k$\textsuperscript{th} constant-Q basis, and $*$ denotes convolution operation. We show the MFSC feature extraction process in continuous time-domain representation ($t$) for similarity with analysis performed in~\cite{anden2014}.  }
    \label{block_diagram}

\end{figure}

We now show similar analysis on CQT to investigate the time and deformation invariance of CQT. From Eq.~\ref{eq:CQT}, the time-shifted version of CQT is given as,

\begin{equation}
    X^{\mathrm{CQT}}[k,n-c]~=~
    \sum_{p=(n-c)-\floor{N_k/2}}^{(n-c)+\floor{N_k/2}}~x(p)a_k^*(p-(n-c)+N_k/2)
    \label{CQT}
\end{equation}
\noindent where, $k$ and $n$ are the frequency and time bins respectively, $N_k$ is the length of $k$\textsuperscript{th} basis function, $c$ is the time shift, $X^{\mathrm{CQT}}$ is the corresponding CQT coefficient, and,

\begin{equation}
    a_k^*(n-c)~=~\frac{1}{N_k}W\left(\frac{n-c}{N_k} \right)exp\left[-i2\pi 
    (n-c)\frac{f_k}{f_s}\right].
\end{equation}

\noindent Therefore, the relation between CQT of time shifted and original version of signal $x$ is given as,

\begin{equation}
    X^{\mathrm{CQT}}[k,n-c]~=~exp\left[-i2\pi c \frac{f_k}{f_s}\right]X^{\mathrm{CQT}}[k,n].
\end{equation}

\noindent Again applying modulus on both sides,

\begin{equation}
    \left|X^{\mathrm{CQT}}[k,n-c]\right|~=~\left|X^{\mathrm{CQT}}[k,n]\right|.
\end{equation}

\begin{figure}[t!]
    \centering
    \includegraphics[scale=0.45]{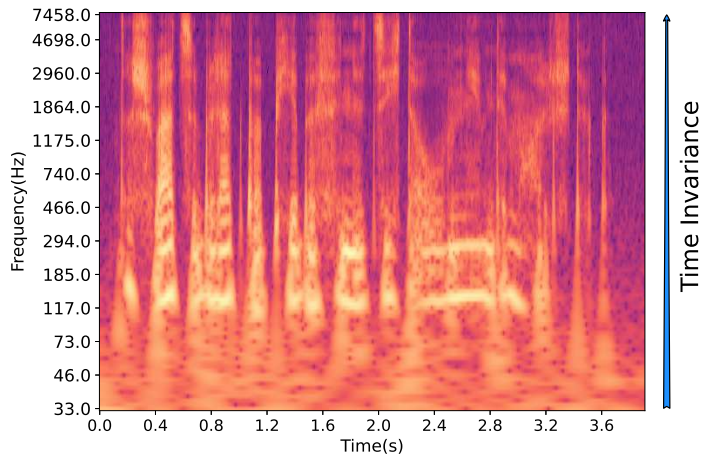}
    \caption{The frequency varying time invariance in CQT. Due to higher value of $N_k$, smearing in time domain is visible at low frequencies.}
    \label{ti_fig}
\end{figure}

\noindent We observe that the modulus of CQT exhibits the same time-shift invariance as observed in STFT. However, in CQT, the time window over which the transform is computed is \emph{not fixed}. The time window duration varies with the relation $N_k=Q\frac{f_s}{f_k}$. Therefore, the time-invariance of CQT also varies with the frequency bin $k$. Since the value of $N_k$ is higher for small values of $k$ and vice-versa, the time-invariance varies from high to low towards high frequencies in the CQT representation. The same is also evident from the smearing of frequency components observed in Fig.~\ref{ti_fig}. This is in contrast with the fixed time-invariance in MFSC, defined by the duration of window function $\phi(t)$. This property of CQT should help provide more time-invariance to the emotion relevant characteristic of speech, i.e., the pitch frequency~\cite{williams1972emotions, banse1996acoustic}. With higher invariance the effect of emotion irrelevant variations in pitch, e.g., variations due to different speaking styles, contexts, etc., can be reduced, hence improving the complete emotion representation. At the same time, the quick variations in time that takes place at high frequency can also be simultaneously captured due to higher time-resolution. This property is more useful for emotions with comparatively more energy at high frequencies, for example, \emph{Angry} and \emph{Happy}. The Eq.~\ref{mfsc_final} equivalent of CQT can be given as,

\begin{equation}
    X^{\mathrm{CQT}}~=~|x*\psi_{\lambda}|(t).
\end{equation}

\begin{figure*}[t!]
    \centering
    \begin{subfigure}[t]{0.5\textwidth}
        \hbox{\hspace{-0.25cm}\includegraphics[scale=0.55]{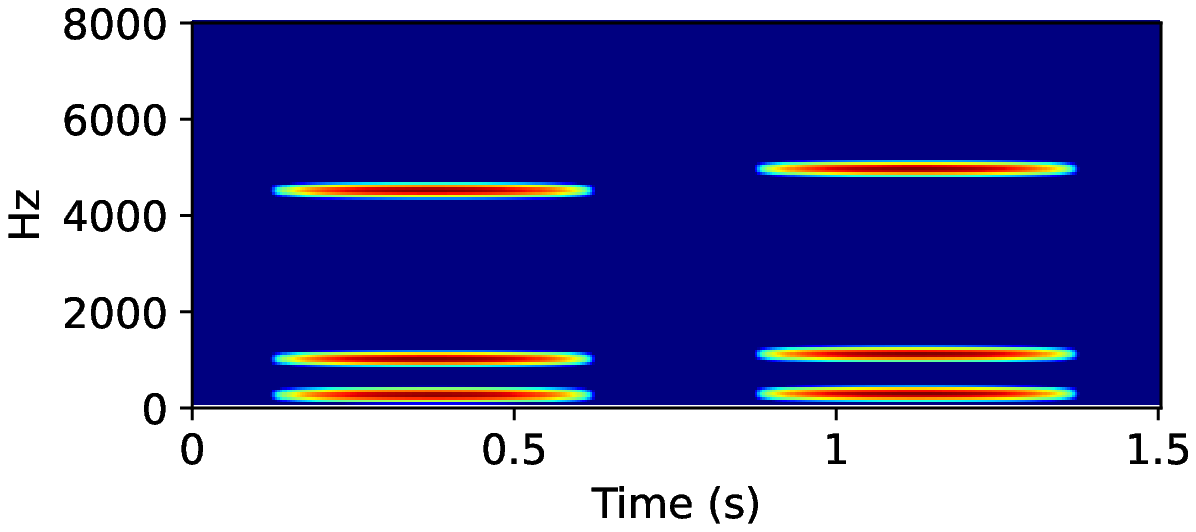}}
        \caption{STFT}
    \end{subfigure}%
    \begin{subfigure}[t]{0.55\textwidth}
        \includegraphics[scale=0.55]{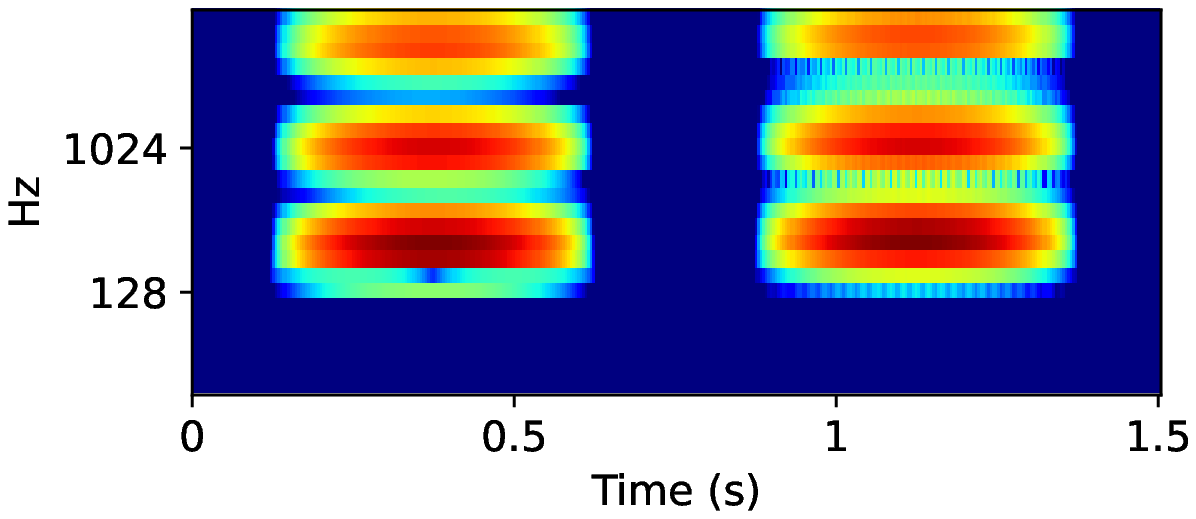}
        \caption{CQT}
    \end{subfigure}%
    \vspace{0.75cm}
    \begin{subfigure}[t]{0.6\textwidth}
        \hbox{\hspace{0.35cm}\includegraphics[scale=0.55]{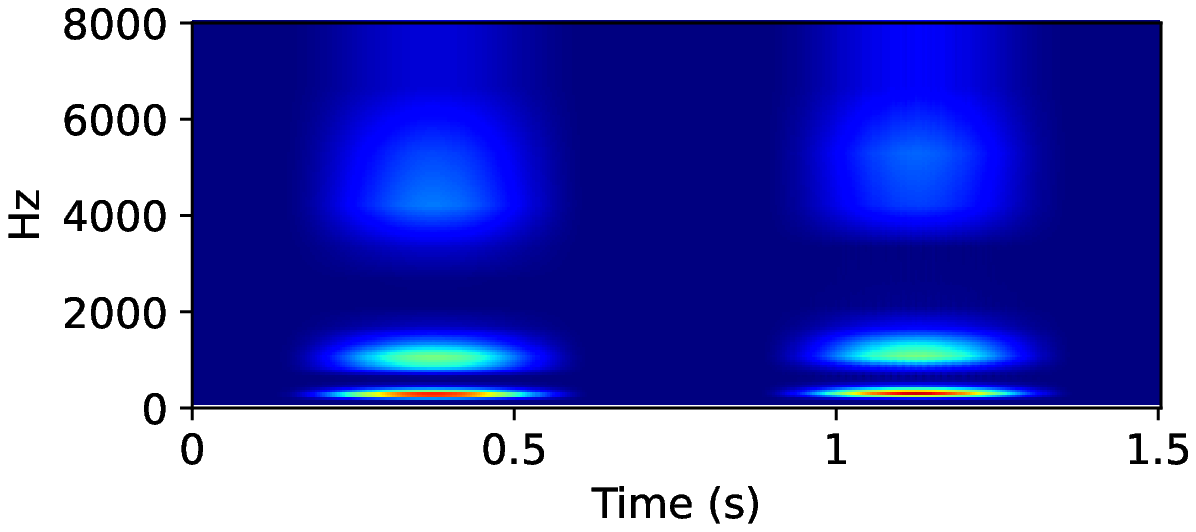}}
        \caption{CQT with linear frequency scale}
    \end{subfigure}
    \caption{Comparison between deformation stability of STFT, CQT, and CQT with linear frequency scale. The figures include frequency response of a synthetic signal $x(t)$ consisting of three tones of $0.5s$ duration (on left) and its deformed counter-part $x'(t) = x((1-\epsilon)t)$ (on right). $x(t)$ is made by adding three tones of $250$~Hz, $1000$~Hz, and $4500$~Hz frequency. In a), harmonics with high frequency get more severely affected by the distortion. In b), due to filters with high bandwidth, the harmonics in the original ($x(t)$) and deformed signals ($x'(t)$) overlap significantly, hence reducing the effect of distortion at high frequencies. In c), we show the linear frequency scale response of $x(t)$ and $x'(t)$ with CQT. The varying bandwidth of filters leads the CQT representation to inherent stability against time-warp deformation.}
    \label{deform_stable}
\end{figure*}

Hence, the time-frequency representation of CQT is the modulus of convolution of signal $x$ with different filters ($\psi_{\lambda}$) in CQT filterbank. The modulus term provides time-invariance to filterbank convolution output. The middle column of Fig.~\ref{block_diagram} shows different steps in CQT computation. Regarding stability to time-warping deformations, CQT also employs varying bandwidth filters with same effect as averaging across different frequency bands. This introduces similar deformation stability as done by mel-filter bank in MFSC. Figure~\ref{deform_stable} explains the effects of deformations on STFT and CQT and the corresponding stability in CQT. The stability removes the variations appearing due to different speaking styles, e.g., slower or faster speaking rate of different individuals.

\renewcommand{\thesubfigure}{\alph{subfigure}}

\subsection{Comparison between CQT and CWT} 
CWT also provides a time-frequency representation generated from a constant-Q filterbank~\cite{vetterliwavelet}. However, the CWT representation is redundant as it does not include downsampling of band-pass filter responses. Also, due to its larger size, it is difficult to consider raw CWT as input to CNN. To alleviate this, our employed implementation includes CWT computation followed by framing and summation of the coefficients. The Eq.~\ref{mfsc_final} equivalent of CWT computation is given as,

\begin{equation}
    X~=~x*\psi_\lambda (t)
    \label{cwt_desc}
\end{equation}

\noindent where, $\psi_\lambda$ is band-pass wavelet basis with support at $\lambda$. 
The framing and averaging performed on computed CWT can also be given as low-pass filtering of Eq.~\ref{cwt_desc}. Hence,

\begin{equation}
    X^{\mathrm{CWT}}~=~|x*\psi_\lambda|*\phi(t),
    \label{cwt_desc2}
\end{equation}

\noindent where, $\phi$ is the framing window and $X^{\mathrm{CWT}}$ is the corresponding CWT coefficient. The modulus is applied to again remove the phase component and obtain time-invariant response. The right most column of Fig.~\ref{block_diagram} describes the computation of CWT. As the filters used in CWT closely follow the structure of CQT filters, Eq.~\ref{cwt_desc} also provides varying time-invariance and higher low-frequency resolution. However, the averaging in Eq.~\ref{cwt_desc2} fixes the extent of time-invariance, similar to MFSC. In our experiments, CWT's frame duration is also kept equal to $20$~ms following MFSC. Therefore, the time invariance in $X^{\mathrm{CWT}}$ varies for low-frequency bins, i.e., for filters with time span greater than $20~$ms, and remain fixed to $20~$ms for filters with span less than $20~$ms. Eq.~\ref{cwt_desc2} closely follows the time-domain MFSC (Eq.~\ref{mfsc_final}) and the first layer scattering domain coefficients explained in~\cite{anden2014} for 1D signals. The comparison of CQT with the CWT time-frequency representation used in this work would help to analyze the importance of varying time-invariance and improved low-frequency resolution, as compared to standard MFSC. This is further consolidated with the experiments that follow in later sections.

\begin{figure*}[t!]
    \centering
    \begin{subfigure}{0.6\textwidth}
        \includegraphics[scale=0.325]{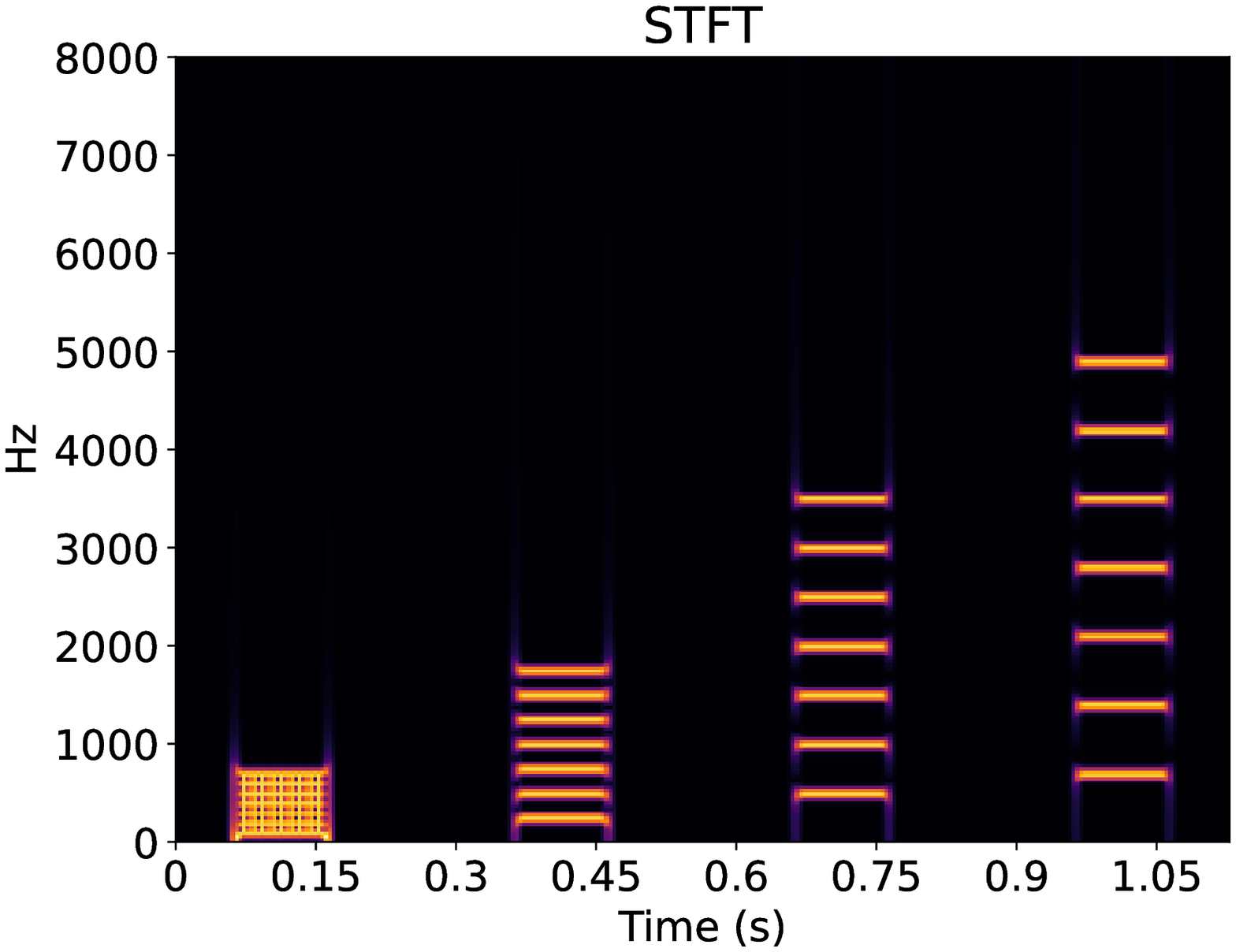}
    \end{subfigure}%
    \begin{subfigure}{0.6\textwidth}
        \includegraphics[scale=0.325]{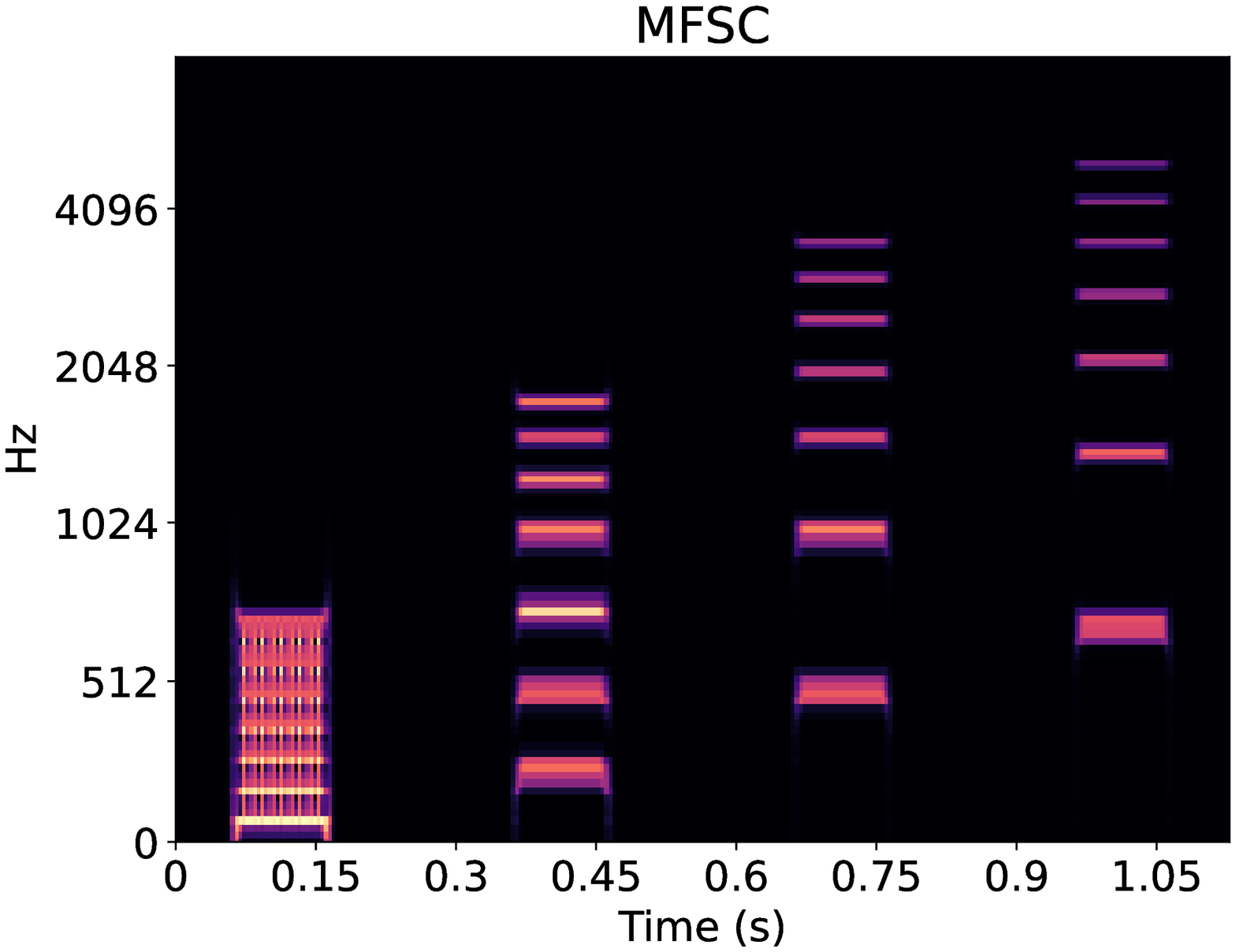}
    \end{subfigure}
    \begin{subfigure}{0.6\textwidth}
        \hbox{\hspace{0.8cm}\includegraphics[scale=0.325]{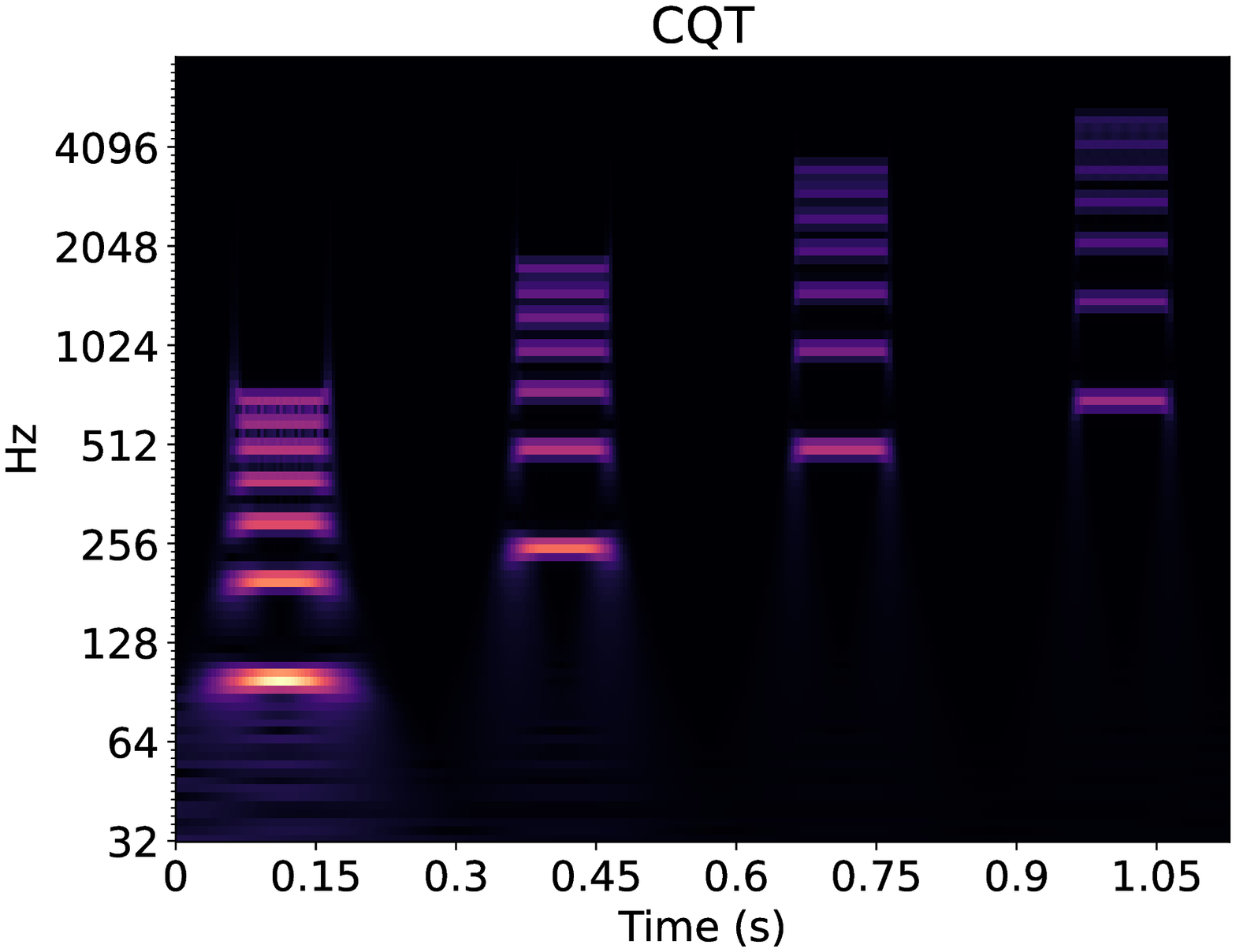}}
    \end{subfigure}
    \caption{Time-frequency representation of signal consisting of $100$~Hz, $250$~Hz, $500$~Hz, and $700$~Hz frequency tones with corresponding first six harmonics. The complete signal is made by concatenating the sub-sequences (tones with corresponding harmonics) with each other. In STFT, the distance between harmonics increases with increase in fundamental frequency (pitch). For log-frequency based scale (mel-scale and constant-Q scale), the distance between harmonics should remain invariant to pitch frequency. In mel-scale, due to the decadic logarithm scale, the distance varies for low pitch values. For CQT, distance remains unchanged for every pitch value. }
    \label{pitch_sep}
\end{figure*}

\begin{figure}[ht]

    \begin{minipage}{0.5\textwidth}
    \hbox{\hspace{-1.8cm}\includegraphics[scale=0.195]{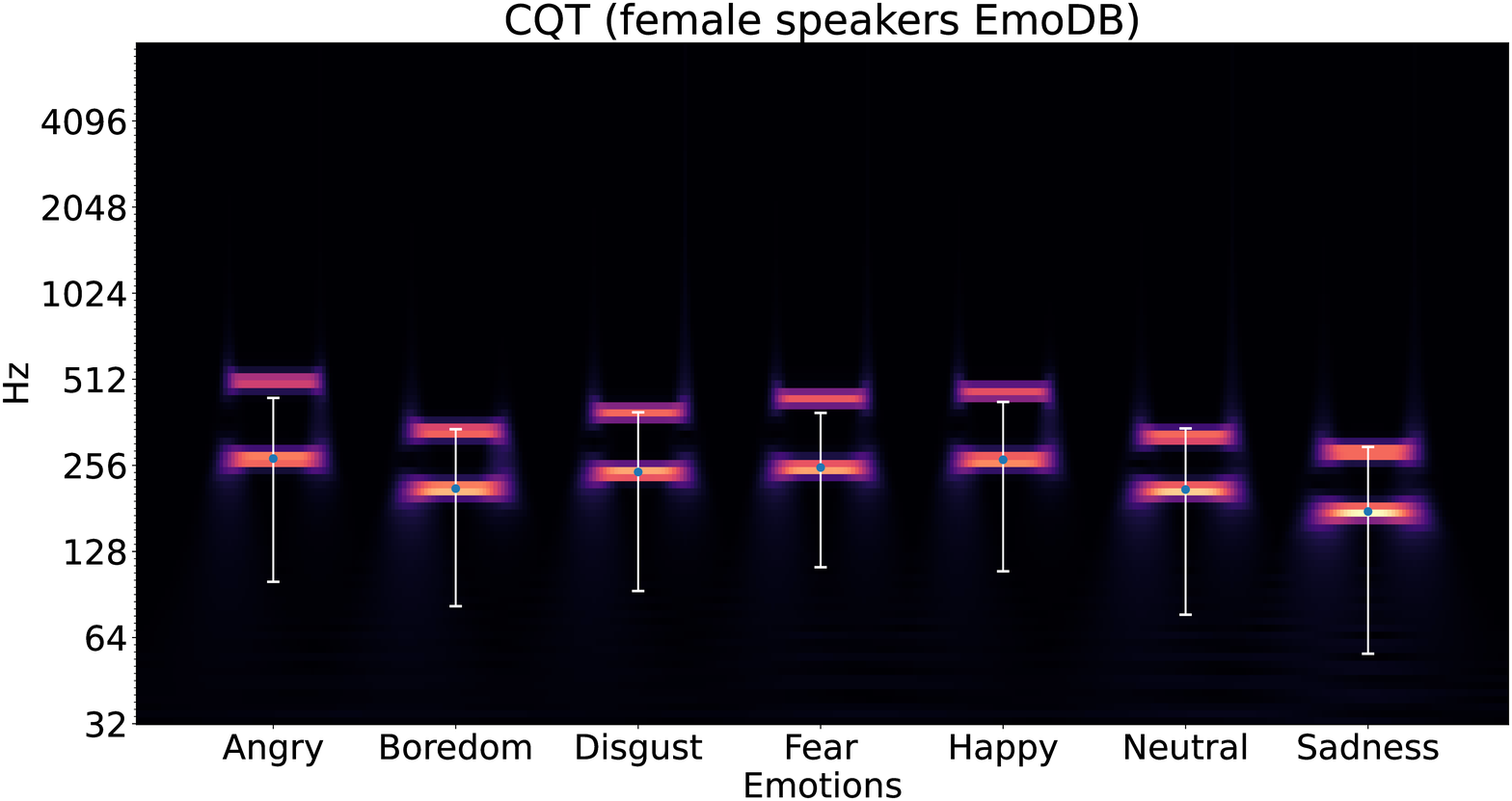}}
    \end{minipage}%
    \begin{minipage}{0.5\textwidth}
    \hspace{0.05cm}
    \includegraphics[scale=0.195]{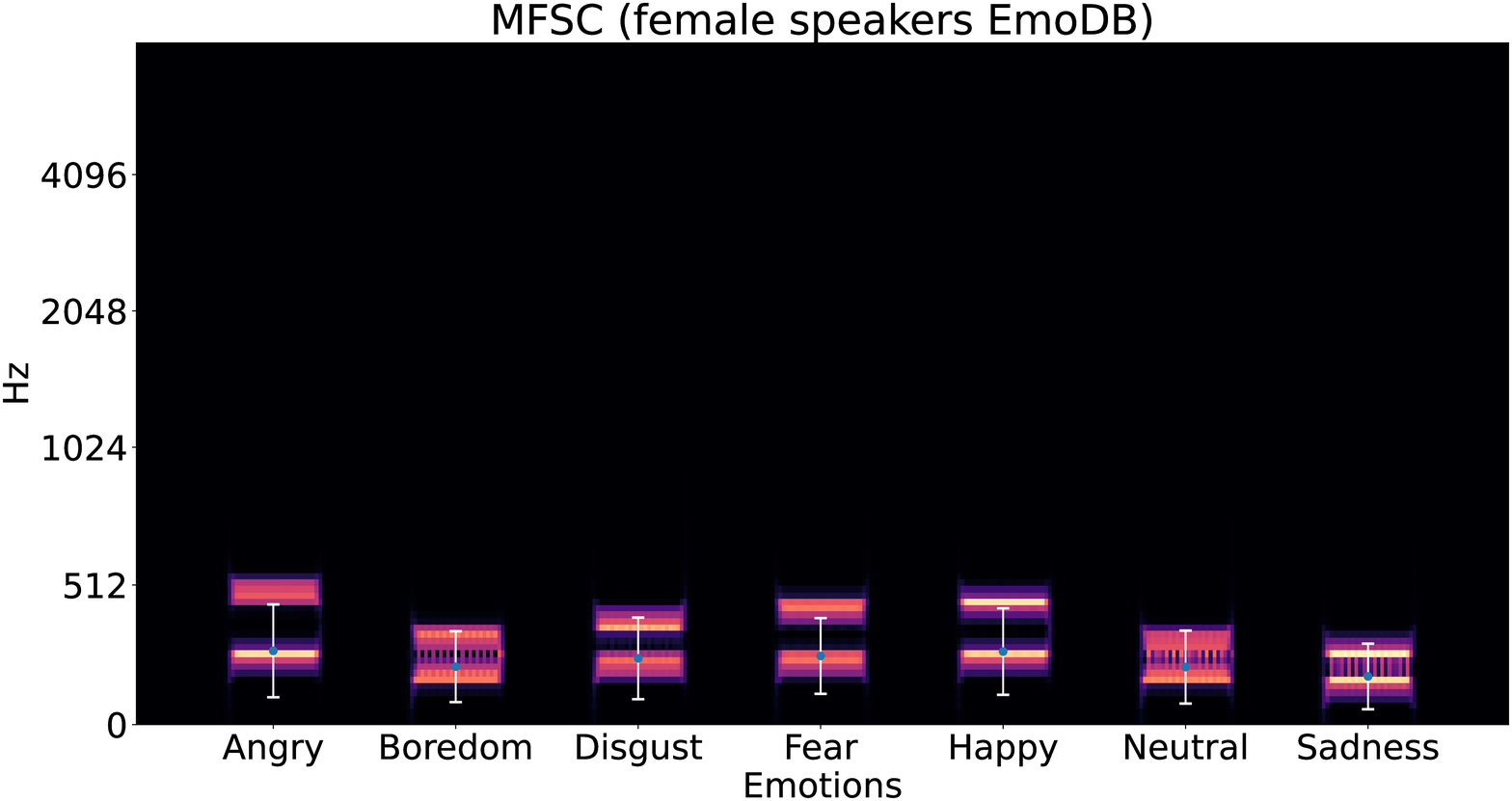}
    \end{minipage}
    
    \begin{minipage}{0.5\textwidth}
    \vspace{0.2cm}
    \hbox{\hspace{-1.8cm}\includegraphics[scale=0.195]{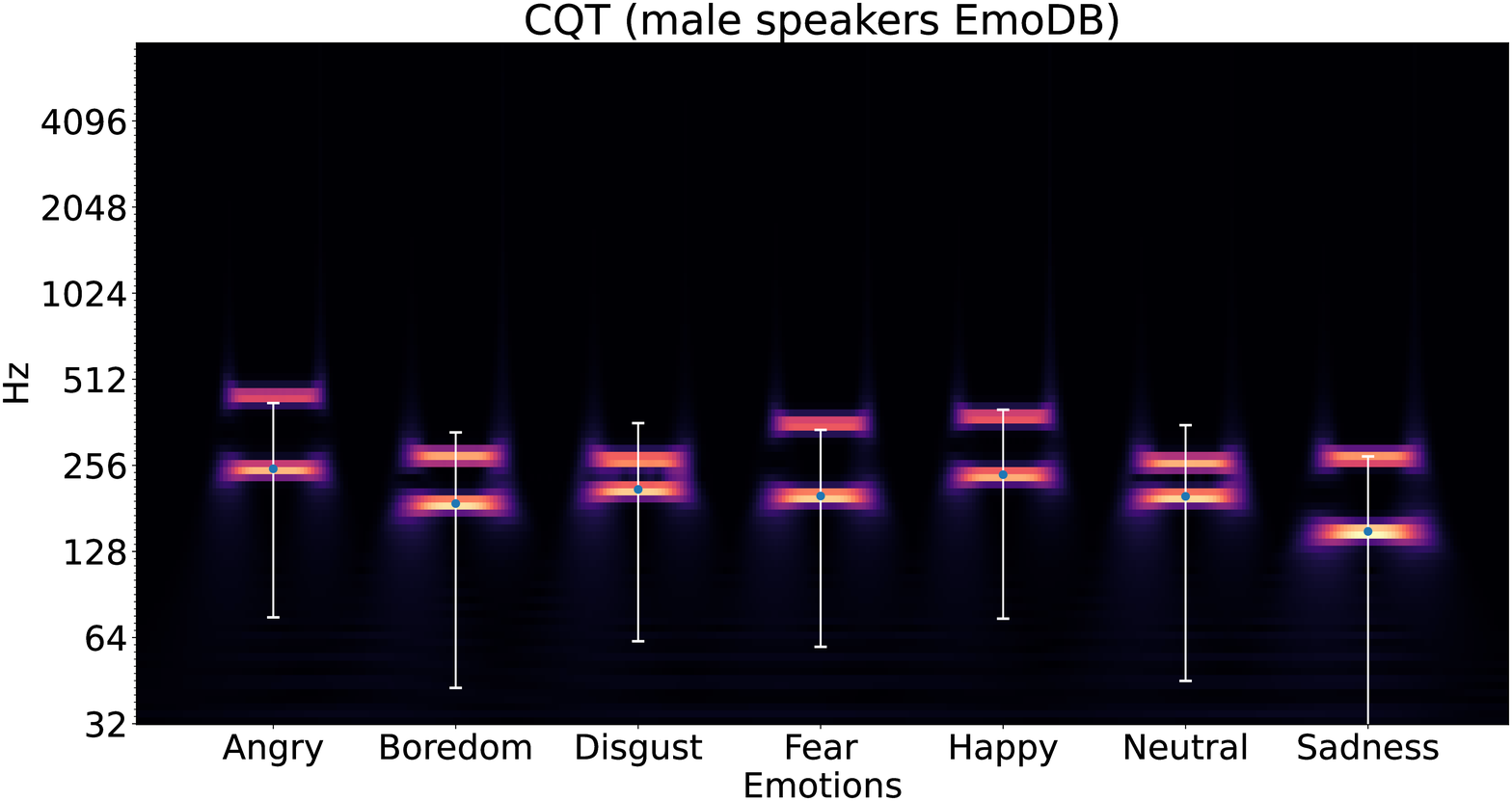}}
    \end{minipage}%
    \begin{minipage}{0.5\textwidth}
    \vspace{0.2cm}
    \hspace{0.05cm}
    \includegraphics[scale=0.195]{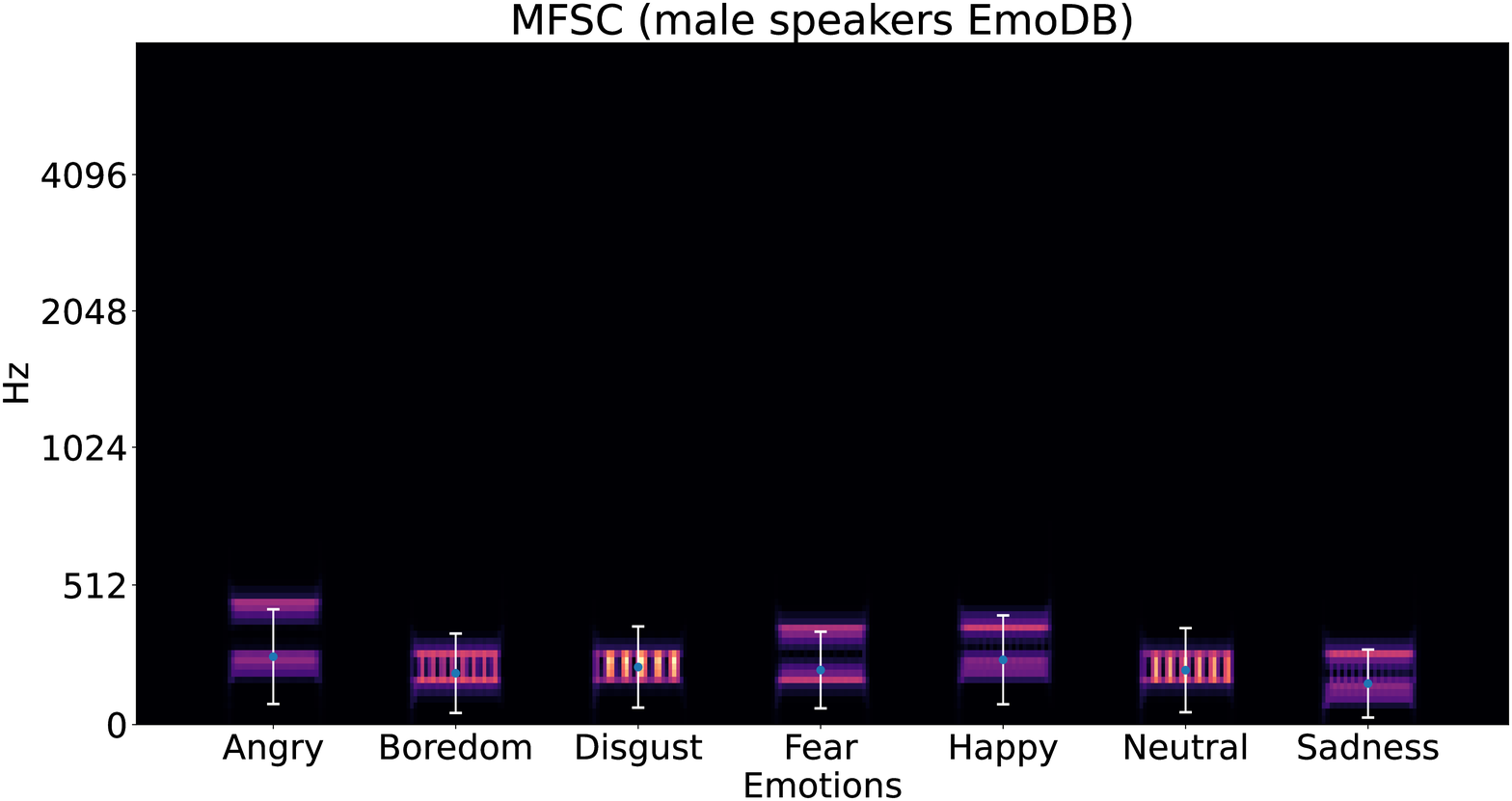}
    \end{minipage}
    \caption{Pitch and first pitch harmonic of Female and Male speakers for various emotions in EmoDB. In the figure, for every emotion, the lower stripe shows the mean pitch averaged over all utterances in EmoDB. The upper stripe shows the mean first pitch harmonic averaged over EmoDB. The error bars show the standard deviation of pitch values for different emotions. Due to higher low-frequency resolution, CQT can better resolve and differentiate between pitch and its first harmonic across both female and male speakers.}
    \label{pitch_harmonic_sep}
\end{figure}

\subsection{Frequency domain comparison of mel and constant-Q filterbanks}
\label{freq_domain_comp}

The major difference between the MFSC and CQT time-frequency representation in time domain appears in the form of fixed and varying time invariance across different frequency bins. In frequency domain, the major difference exists in the center frequencies of filters in the filterbank. Mel-scale follows a decadic logarithm scale, whereas CQT follows a binary logarithm scale. This leads to higher low-frequency resolution in CQT as compared to MFSC. Another property of the log-frequency is that the distance between pitch harmonics is invariant to pitch frequency. This is contrast with in linear frequency representation (e.g., spectrogram) where the harmonic distance reduces with reduced pitch frequency. Figure~\ref{pitch_sep} describes this phenomenon. This also helps CQT (or constant-Q response, in general) in better resolution of pitch and its harmonics as compared to STFT or MFSC. For further analysis, we compute the mean and standard deviation of pitch and first pitch harmonic, averaged across every utterance for different emotions of male and female speakers in EmoDB database. We then generate tones corresponding to average pitch and its first harmonic frequency and compute the CQT and MFSC representation of the tones. Figure~\ref{pitch_harmonic_sep} shows the obtained representations. We observe that CQT better resolves the pitch and its first harmonic as compared to MFSC for both male and female speakers. MFSC shows overlapping between pitch and its harmonic for emotions with low mean pitch frequency (\emph{Boredom}, \emph{Neutral} and \emph{Sadness}). However, CQT clearly differentiates pitch and its harmonics for every emotion in EmoDB. Also, for \emph{Disgust}, \emph{Neutral} and \emph{Sad} emotions, the separation in CQT representation of females is higher than that in males. This is because of the naturally higher pitch of female speakers as compared to that of the males. The error bars show the standard deviation of pitch for various emotions in EmoDB. Same value of standard deviation show different dynamic range over frequency bins in CQT and MFSC. This emphasizes the difference in low-frequency resolution in MFSC and CQT and also the superior pitch resolution in CQT. 

The studies in~\cite{LU} and~\cite{SARANGI} suggest that a filterbank structure with high resolution (dense filters) on mid and high frequency regions are beneficial for speaker recognition. In contrast to this, the dense filter arrangement at low-frequency in CQT and CWT should provide speaker complimentary information. Therefore, constant-Q representation should also remain invariant to speaker information compared to MFSC (because of more low-frequency resolution) facilitating SER.

\section{Experimental setup}

\subsection{Neural network architectures}

\label{nn_arch}
In this subsection, we briefly review the different deep neural network architectures that were employed to evaluate SER performance of features. Our choice of these architectures was inspired by the success of techniques such as 1D and 2D convolutions, LSTM, attention mechanism, squeeze and excitation module, Res2Net module, etc. in different speech processing domains~\cite{huang2017characterizing, tang2018end, schullerlstm, ctnet, dawalatabad2021ecapa, desplanques2020ecapa}.

\begin{table}[ht]
\renewcommand{\arraystretch}{1.1}
\caption{Parameters of TDNN, Conv2D, and Conv2D-LSTM architecture used in this work.}
\begin{adjustbox}{width=0.75\textwidth,center}
\begin{tabular}{cccccc}
\hline
\textbf{Architecture} & \textbf{Layers} & \textbf{No. of Filters} & \begin{tabular}[c]{@{}c@{}}\textbf{Height} \\ \textbf{(Frequency)} \end{tabular} & \begin{tabular}[c]{@{}c@{}} \textbf{Length} \\ \textbf{(Time)} \end{tabular} & \textbf{Dilation} \\ 
\hline \hline
\multirow{7}{*}{TDNN} & 1D Conv & 32 & 5 & 5 & 1 \\
 & 1D Conv & 32 & 3 & 3 & 2 \\
 & 1D Conv & 32 & 3 & 3 & 3 \\
 & 1D Conv & 64 & 1 & 1 & 1 \\
 & \begin{tabular}[c]{@{}c@{}}Mean \& Standard\\ Deviation Pool\end{tabular} & - & - & - & - \\
 & Fully Connected & 64 & - & - & - \\
 & Softmax & \#Classes & - & - & - \\ \hline 
\multirow{9}{*}{Conv2D} & 2D Conv & 32 & 5 & 5 & 1 \\
 & MaxPool & - & 2 & 1 & 1 \\
 & 2D Conv & 32 & 3 & 3 & 1 \\
 & 2D Conv & 32 & 3 & 3 & 1 \\
 & 2D Conv & 64 & 1 & 1 & 1 \\
 & \begin{tabular}[c]{@{}c@{}}Mean \& Standard\\ Deviation Pool\\ (over time)\end{tabular} & - & - & - & - \\
 & Flatten & - & - & - & - \\
 & Fully Connected & 64 & - & - & - \\
 & Softmax & \#Classes & - & - & - \\ \hline
\multirow{14}{*}{Conv2D-LSTM} & 2D Conv & 64 & 3 & 5 & 1 \\
 & MaxPool & - & 2 & 2 & 1 \\
 & 2D Conv & 64 & 3 & 5 & 1 \\
 & MaxPool & - & 2 & 2 & 1 \\
 & 2D Conv & 64 & 3 & 5 & 1 \\
 & MaxPool & - & 2 & 2 & 1 \\
 & 2D Conv & 64 & 3 & 5 & 1 \\
 & MaxPool & - & 2 & 2 & 1 \\
 & 2D Conv & 64 & 3 & 5 & 1 \\
 & 2D Conv & 64 & 3 & 5 & 1 \\
 & \begin{tabular}[c]{@{}c@{}}Mean Pool\\ (over frequency)\end{tabular} & - & - & - & - \\
 & BLSTM & 64 & - & - & - \\
 & Fully Connected & 64 & - & - & - \\
 & Softmax & \#Classes & - & - &  \\ \hline
\end{tabular}
\end{adjustbox}
\label{param_tab}
\end{table}

\subsubsection{TDNN architecture}
As a 1D convolutional network, a TDNN architecture, shown in Table~\ref{param_tab}, was used for SER with the time-frequency representations as inputs. The TDNN structure was inspired by the \emph{x-vector} systems developed for speaker verification~\cite{snyder2018x}. A TDNN structure processes frame-level information in the convolutional layer by applying different dilation values at different layers to efficiently capture the temporal spread of information. This is followed by a statistics pooling layer which aggregates the processed information in the temporal dimension to generate a segment level representation. The segment-level features are then fed to FC layers and a softmax layer for final classification. The TDNN structure used in this work employed an end-to-end classification system, in contrast with the structure used in~\cite{snyder2018x}, which extracts segment-level speaker embeddings from the statistics pooling layers for classification with a different classifier.

\subsubsection{Conv2D architecture}
Table~\ref{param_tab} also shows the architecture of a 2-dimensional CNN or Conv2D network used for SER classification. Conv2D network involves convolution of the input feature matrix with a kernel which allows movement in both time and frequency dimensions of input time-frequency representation. This helps in capturing the feature correlation across both time and frequency dimensions. As the emotion information is known to remain temporally spread throughout the utterance, 2D convolution windows with different kernel sizes helps in information extraction across different time scales. Similar to the TDNN architecture, our Conv2D architecture also applied temporal mean and standard deviation pooling on the output of the final convolution layer. This operation again aggregates the features extracted by convolutional layers across time dimension to obtain a vector representation of the input segment. The temporal pooling is followed by FC and Softmax layers for classification. We used the Conv2D network for end-to-end classification, with speech features at the input and probable emotion class at the output. We kept the parameters, i.e., no.~of filters, kernel sizes, and number of layers in Conv2D same as used in TDNN architecture for comparison. We also placed a max-pooling layer after the first convolutional layer to reduce parameters in Conv2D architecture and prevent overfitting.

\subsubsection{MERC2020 baseline architecture}
To compare SER performance across different architectures, we used the speech emotion recognition model proposed in multimodal emotion recognition competition 2020 (MERC 2020). The model contains three Conv2D layers, one LSTM layer followed by attention pooling and a dense layer. We used the implementation provided by the organizers of the MERC2020\footnote{\url{https://github.com/ki4ai-skc/merc2020}}.

\subsubsection{Attention-based LSTM}
We also evaluated the performance of different features with attention-based LSTM model proposed in~\cite{schullerlstm}. The model includes a modified LSTM, in which the forget gate is replaced with an attention mechanism called attention gate. The attention gate provides increased focus on the most emotion-relevant parts of the input time-frequency representation. Moreover, this modification decreases the number of trainable parameters in every attention-based LSTM block, causing a reduction in the model training time~\cite{schullerlstm}. The output of attention-based LSTM is fed to two separate attention layers, one focusing on time and the other on the frequency dimension. Hence, the complete architecture consists of two layers of modified LSTM units (attention-based LSTM blocks) followed by parallelly placed time and frequency attention layers. The activations from attention layers are concatenated and fed to two FC layers and softmax for final classification. We chose the same parameter values for different layers as used in the original paper.

\subsubsection{Transformer encoder model}
We employed a transformer encoder model proposed in~\cite{ctnet} to compare with the state-of-the-art deep learning architectures. In~\cite{ctnet}, information from various modalities (speech, speaker, and text) was extracted using encoder blocks, and the output was concatenated for SER. Since our approach includes only speech modality, we chose the speech signal's transformer encoder block from~\cite{ctnet} in our experiments. The model consists of a Conv1D layer followed by a combination of multi-head attention and linear layers, constituting the encoder block. The output of the encoder block is then fed to a frame-level (or time-level) attention pooling layer followed by an FC and softmax. The parameter values of the neural network layers were the same as used in the original paper.

\subsubsection{ECAPA-TDNN}
The emphasized channel attention, propagation, and aggregation in time-delay neural network (ECAPA-TDNN) proposed in~\cite{desplanques2020ecapa} introduces multiple enhancements over the standard x-vector TDNN architecture~\cite{snyder2018x}. These include using multiple 1D Res2Net modules, squeeze and excitation blocks, and channel-dependent time-frame attention. The ECAPA-TDNN architecture is found useful for speaker recognition and speaker diarization tasks~\cite{dawalatabad2021ecapa}. In this work, we used the implementation of ECAPA-TDNN provided in SpeechBrain\footnote{\url{https://speechbrain.github.io/}} Python toolkit without any change in parameter configuration.

\subsubsection{Conv2D-LSTM}
In~\cite{xuanjihe2018}, an attention-based convolutional recurrent neural network (ACRNN) was proposed for SER. Inspired by this, we designed an architecture consisting of only CNN and LSTM layers. We removed the attention layer from this architecture to analyze the effect of the temporal memory-based recurrent layer on CNN learned features and compare it with plain Conv2D architecture. Also, the MERC2020 model already included a combination of LSTM and attention with convolution layers. Table~\ref{param_tab} describes our employed Conv2D-LSTM architecture. An LSTM contains a memory element that can accumulate the required information across several time frames. Using LSTM on CNN activations helps extract emotion-relevant temporal information from translation invariant and downsampled feature representations provided by CNN. Since the last time-frame output of the LSTM contains the most abundant emotion information~\cite{schullerlstm}, we fed only this output to the final FC and softmax layers and discarded the remaining time frames. This architecture also helped to compare the temporal aggregation of emotion information of LSTM with that of attention layers in Attention-based LSTM, Transformer encoder, and ECAPA-TDNN architectures. 

For the above mentioned deep networks, we used \emph{Keras}\footnote{\url{https://keras.io/}} deep learning library for Conv2D, TDNN, and Conv2D-LSTM architecture and \emph{PyTorch}\footnote{\url{https://pytorch.org/}} deep learning library for the remaining selected state-of-the-art architectures.

\subsection{Databases}
\label{databases}
We used four different databases for analysis and evaluation of features. They are freely available and widely used in SER. Performance of selected features on these corpora also facilitates comparison with similar SER methods. Table~\ref{tab2} summarizes the different SER databases used in our experiments.

\subsubsection{Berlin emotion database (EmoDB)}
Berlin emotion database (EmoDB)~\cite{burkhardt2005database} is one of the most widely used database in SER. It includes acted spoken utterances of $10$ professional artists ($5$ female and $5$ male). Ten sentences, emotionally neutral and phonetically rich, are spoken by the actors in German language. The database contains speech recordings of seven different emotions: \textit{Angry, Happy, Fear, Sad, Boredom, Disgust} and \textit{Neutral}. The authenticity of the recorded emotions was evaluated by listening test performed on $20$ subjects. In total, the actors recorded $800$ utterances but only $535$, having more than $80\%$ recognition rate and $60\%$ naturalness, were finally selected. The mean recognition accuracy of emotions in the listening test was $84.3\%$ on the selected $535$ recordings. The diligent recording setup and free availability has led this database to be used in various important works~\cite{gemaps, mao2014learning, zhang2017speech, bitouk2010class, wu2011automatic, wang2015speech, deb2018multiscale, Ntalampiras2012} and hence is used here as well.

\subsubsection{Ryerson audio-visual database of emotional speech and song (RAVDESS)}
The RAVDESS database~\cite{livingstone2018ryerson} contains emotion speech and song samples recorded from $12$ male and $12$ female artists speaking English language. The database contains a total of $7536$ clips with data recorded in three modalities: audio-only, video-only, and audio-video. The audio-only modality contains $1440$ speech utterances from all speakers spoken with eight different emotions (\textit{Happy, Angry, Sad, Neutral, Disgust, Calm, Surprised} and \textit{Fear}) and two intensity levels, strong and normal. Evaluation of recorded clips were performed by $319$ subjects, out of which $247$ evaluated the validity and $72$ provided test-retest reliability of recorded emotions. An average of $60\%$ accuracy was obtained in validity test on recordings of all emotions. Recent design and inclusion of an extensive emotion set with varying intensities make this an important database for SER.

\begin{table*}[t]
\centering

\caption{Summary of SER Databases.}
\begin{adjustbox}{width=1.2\textwidth,center}
\label{tab2}
\begin{tabular}{@{}cccccccc@{}}
\toprule
\textbf{Databases} & \textbf{Type} & \textbf{Speakers} & \textbf{Emotions} & \textbf{\begin{tabular}[c]{@{}c@{}}Sampling \\ Rate\end{tabular}} & \textbf{Total Utterances} & \textbf{\begin{tabular}[c]{@{}c@{}} Language \end{tabular}} \\ \midrule \midrule

\begin{tabular}[c]{@{}c@{}}Berlin Emotion Database (EmoDB) \cite{burkhardt2005database}\end{tabular} & Acted & \begin{tabular}[c]{@{}c@{}}$10$\\ ($5$ Female)\\ ($5$ Male)\end{tabular}  & \begin{tabular}[c]{@{}c@{}}7\\ (Anger, Sad, Boredom,\\ Anxiety/Fear, Happy,\, \\ Disgust and Neutral)\end{tabular} & $16$~kHz & $535$ & German \\
\vspace{0.2cm}
\begin{tabular}[c]{@{}c@{}}Ryerson Audio-Visual Database of \\ Emotional Speech and Song (RAVDESS) \cite{livingstone2018ryerson}\end{tabular} & Induced &\begin{tabular}[c]{@{}c@{}}24\\ (12 Female)\\ ($12$ Male)\end{tabular} & \begin{tabular}[c]{@{}c@{}}8\\ (Calm, Happy, Sad, \\ Anger, Neutral, Fearful, \\ Surprise, and Disgust)\end{tabular} & $48$~kHz & \begin{tabular}[c]{@{}c@{}} $1440$ \\ \end{tabular}& English \\ 

\begin{tabular}[c]{@{}c@{}}Interactive Emotional Dyadic Motion \\ Capture Database (IEMOCAP) \cite{busso2008iemocap} \end{tabular} & Induced & \begin{tabular}[c]{@{}c@{}}10\\ ($5$ Female)\\ ($5$ Male)\end{tabular}  & \begin{tabular}[c]{@{}c@{}}4\\ (Happy, Sad, Anger \\ and Neutral) \end{tabular} & $16$~kHz & \begin{tabular}[c]{@{}c@{}} $4936$  \\ \end{tabular} & English \\ 

\begin{tabular}[c]{@{}c@{}} eNTERFACE '05 \cite{martin2006enterface} \end{tabular} & Induced & \begin{tabular}[c]{@{}c@{}}44\\ (8 Female)\\ ($36$ Male)\end{tabular} & \begin{tabular}[c]{@{}c@{}}6\\ (Happy, Sad, Anger,\\ Fear, Surprise\\ and Disgust)\end{tabular} & $48$~kHz & \begin{tabular}[c]{@{}c@{}} $1293$ \\ \end{tabular} & \begin{tabular}[c]{@{}c@{}} English \\ (Different \\ nationality) \\ \end{tabular} \\ 

\bottomrule
\end{tabular}
\end{adjustbox}
\label{datatable}
\end{table*}

\subsubsection{Interactive Emotional Dyadic Motion Capture Database (IEMOCAP)}
IEMOCAP is an audio-visual database recorded on $10$ professionally trained English speakers ($5$ male and $5$ female) with two recording methods, scripted and improvised~\cite{busso2008iemocap}. This makes the IEMOCAP recordings more natural compared to the two databases mentioned previously. Eight emotions (\textit{Happy, Angry, Sad, Neutral, Fear, Disgust, Excitement} and \textit{Surprise}) were captured over a total of $10039$ recorded samples ($5255$ scripted and $4784$ spontaneous) with an average utterance length of $4$ seconds. Samples were annotated into both discrete and continuous emotion labels by six evaluators. In our work, we used discrete emotion labels from only four classes (\textit{Happy, Angry, Sad} and \textit{Neutral}) of both scripted and improvised recordings because of comparatively sparse speech samples for the remaining emotions and also for better comparison with existing SER literature~\cite{parry2019analysis, ghosh2016representation, issa2020speech, fayek2017evaluating, luo2018investigation}.

\subsubsection{eNTERFACE `05}
eNTERFACE is also an audio-visual database containing recording of six different emotions: \textit{Happy, Sad, Surprise, Anger} and \textit{Fear}, recorded from $44$ different subjects~\cite{martin2006enterface}. Although the subjects were from different nationalities, English was the common language for recording the data. This introduced accent variability on recorded samples, representing the real-world scenario in a better way. To induce emotions, subjects were made to read a short story before recording their reactions to the story on a fixed set of answers. This introduced genuineness into the subject's reactions. We used only the audio modality having a total of $1293$ utterances across all the subjects.

\begin{table*}[t!]
\centering
\caption{Optimized parameter settings for different features ($Q$ = Q-factor of the filter, $f_s$ = sampling frequency, $f_k$ = $k$\textsuperscript{th} frequency bin). The parameters values are taken from experiments performed in~\cite{singh2021non}.}
\hspace{-0.1cm}
\renewcommand{\arraystretch}{1.2}
\label{feattable}
\hspace{-0.35cm}
\resizebox{0.8\columnwidth}{!}{%
\begin{tabular}{cccccc}
\cline{1-5}
  \multicolumn{5}{c}{\begin{tabular}{c}\textbf{Mel-frequency features} \\  \textbf{(MFSC)}\end{tabular}}\\ [0.2cm]
\cline{1-5} \midrule
  Library used & \begin{tabular}[c]{@{}c@{}} Window length \\ (samples) \end{tabular} &
  \begin{tabular}[c]{@{}c@{}} Hop length \\ (samples) \end{tabular} &
  \begin{tabular}[c]{@{}c@{}} No. of \\ FFT points \end{tabular} &
  \begin{tabular}[c]{@{}c@{}} No. of \\ filters \end{tabular}\\[0.2cm] \hline
\emph{LibROSA}(v0.7.1) &  320 & 64 & 512 & 24 \\[0.1cm] 
\cline{1-5} 
  \multicolumn{5}{c}{\begin{tabular}{c}\textbf{Constant-Q transform} \\ \textbf{(CQT)} \end{tabular}}\\ [0.2cm]
\cline{1-5} \midrule
  Library used & \begin{tabular}[c]{@{}c@{}} Window length \\ (samples) \end{tabular} &
  \begin{tabular}[c]{@{}c@{}} Hop length \\ (samples) \end{tabular} &
  \begin{tabular}[c]{@{}c@{}} Bins per \\ octave \end{tabular} &
  \begin{tabular}[c]{@{}c@{}} No. of \\ bins \end{tabular}\\[0.2cm] \hline
\emph{LibROSA}(v0.7.1) & Variable ($Q \frac{fs}{fk}$) & 64 & 3 & 24 \\[0.1cm] 
\cline{1-5}
  \multicolumn{5}{c}{ \begin{tabular}{c}\textbf{Wavelet transform} \\ \textbf{(CWT)} \end{tabular}}\\ [0.2cm]
\cline{1-5} \midrule
  Library used & \begin{tabular}[c]{@{}c@{}} Window length \\ (samples) \end{tabular} &
  \begin{tabular}[c]{@{}c@{}} Hop length \\ (samples) \end{tabular} &
  \begin{tabular}[c]{@{}c@{}} CWT scale values \\ and wavelet type \end{tabular} &
  \begin{tabular}[c]{@{}c@{}} No. of \\ bins \end{tabular}\\[0.2cm] \hline
\emph{PyWavelets}(v1.1.1) & 320 & 64 & \begin{tabular}[c]{@{}c@{}} $2^{\frac{k}{3}}: k \in[3:26]$  \\ Complex Morlet \end{tabular} & 24 \\[0.1cm] \hline
\end{tabular}
}
\end{table*}

\subsection{Training/Testing evaluation}
\label{train_test_eval}
Leave-one-speaker-out (LOSO) cross-validation strategy was employed for evaluation of features with DNN classifiers. The databases were divided into training, validation, and test partitions such that the training and validation groups contained sets of disjoint speakers with one left-out speaker kept for testing. The validation set contained speech utterances of only two speakers. Hence, the number of training-validation-testing sets were same as the total number of speakers for every database. According to the literature, speaker-dependent SER generally fairs better than speaker-independent SER~\cite{schuller2005speaker}. However, using speaker-independent sets from the database eliminates the chances of the trained classifier being biased towards a set of speakers, and also simulates the practical/real-world scenario. We used LOSO even for RAVDESS and eNTERFACE databases which have higher number of speakers ($>10$), unlike the leave-one-speaker-group-out (LOSGO) method used in other works, e.g., \cite{zhang2017speech, schuller2009}.

Although LOSO cross-validation is computationally extensive, we can safely ignore the increase in complexity due to the small sizes of available SER databases. Also, using only one speaker for testing provides the advantage of more data samples in training and validation, which is essential for small databases. 

We used an energy-based speech activity detector to remove silence parts of speech utterances before feature extraction. We also employed cepstral mean variance normalization on features before providing them to the classifiers. To increase the size of available training data, we employed five-fold data augmentation using additive and reverberation noises following x-vector extraction recipe~\cite{snyder2018x} used in \emph{Kaldi}\footnote{\url{https://github.com/kaldi-asr/kaldi/tree/master/egs/voxceleb/v2}} toolkit. All available databases were downsampled to $16$~kHz before feature extraction.

\subsection{Feature evaluation}

\begin{figure*}[t!]
    \centering
    \includegraphics[width=13cm, height=5cm]{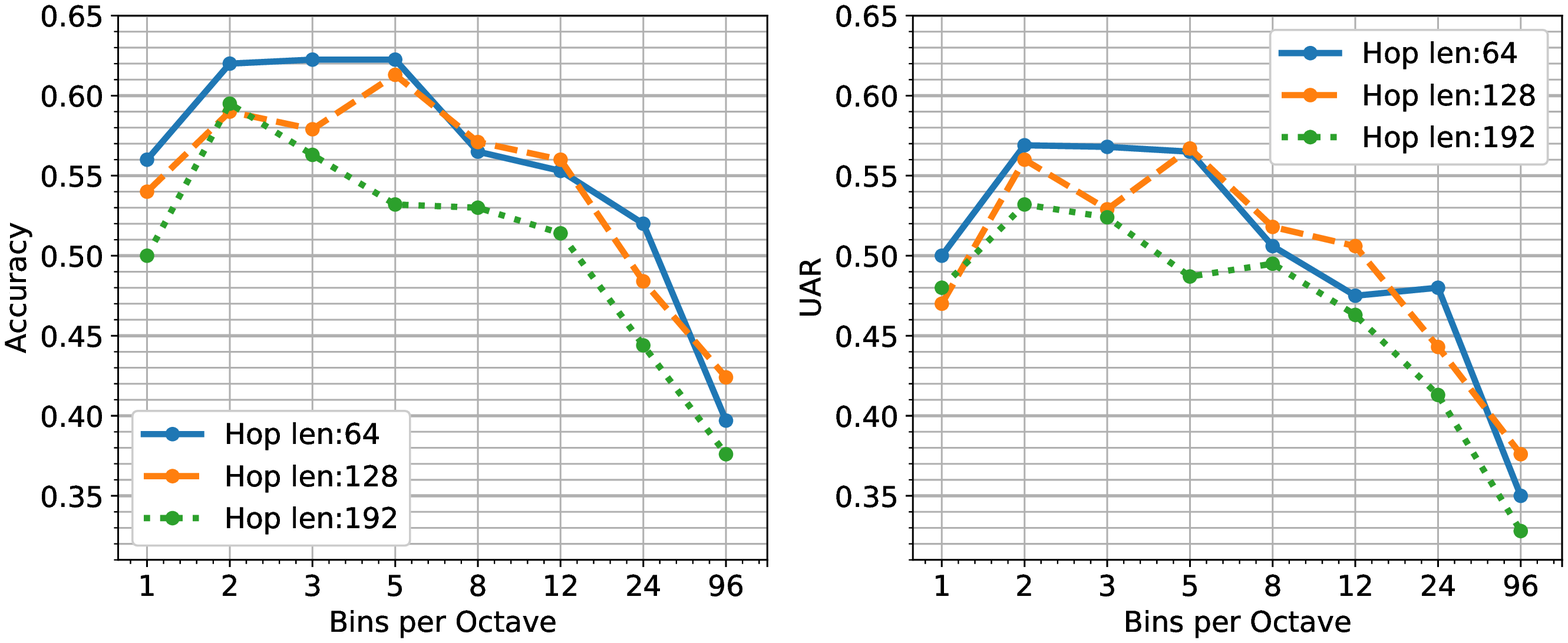}
    \caption{CQT parameter comparison on EmoDB database. The left subplot shows the change in accuracy with different values of frequency bins per octave. Similarly, the right subplot shows change in UAR with frequency bins per octaves. The total number of octaves are kept fixed at 8. Figure taken from~\cite{singh2021non}.}
    \label{CQT_param_comp}
\end{figure*}

Table~\ref{feattable} shows the values of the arguments of the built-in feature extraction functions. \emph{LibROSA}\footnote{\url{https://librosa.github.io/}} \textit{Python library} was used for CQT and MFSC while \emph{PyWavelets}\footnote{\url{https://github.com/PyWavelets/pywt}} was employed for CWT feature extraction in this work. The parameter values were inspired by our preliminary study, where we performed the optimization of bins per octave and hop length over $8$ octaves on EmoDB database~\cite{singh2021non}. For a fair comparison, similar optimization on MFSC and CWT is employed in this work. We used $24$ mel-filter bank with $64$ samples hop across different time frames. This corresponds to the optimized CQT parameters values which provided the best results ($3$ bins per octave with a total $8$ octaves, refer Fig~\ref{CQT_param_comp}). In CWT, we used scale values obtained from the expression $2^{\frac{k}{3}}$ where $k$ varies from $3$ to $26$, to obtain the same number of frequency bins as in CQT. The $3$ in $2^{\frac{k}{3}}$ corresponds to the voices per octave with number of octaves again fixed to $8$. We chose default values for the remaining input parameters of CQT and MFSC functions in \emph{LibROSA} and CWT in \emph{PyWavelets}. \\
 
\subsection{Classifier evaluation}
 
For training of DNN architectures, we used non-overlapping segments of input time-frequency representation of length $100$ frames. However, during testing, we used the utterances' complete duration. These settings improved the final recognition accuracy compared to the scenario where the test data was truncated. We used a learning rate value of $0.001$ with a batch size of $64$. The networks were trained for $50$ epochs, and the model with the highest validation unweighted average recall (UAR) was used during inference. In TDNN, Conv2D, and Conv2D-LSTM architectures, a drop-out value of $0.3$ was used in fully connected layers. ReLU activation was used in every layer except for the final softmax layer.

\begin{figure}
\centering
    \begin{minipage}{\textwidth}
    \subcaptionbox{t-SNE plot of CQT embeddings}
    {\hbox{\hspace{-1.5cm}\includegraphics[scale=0.425]{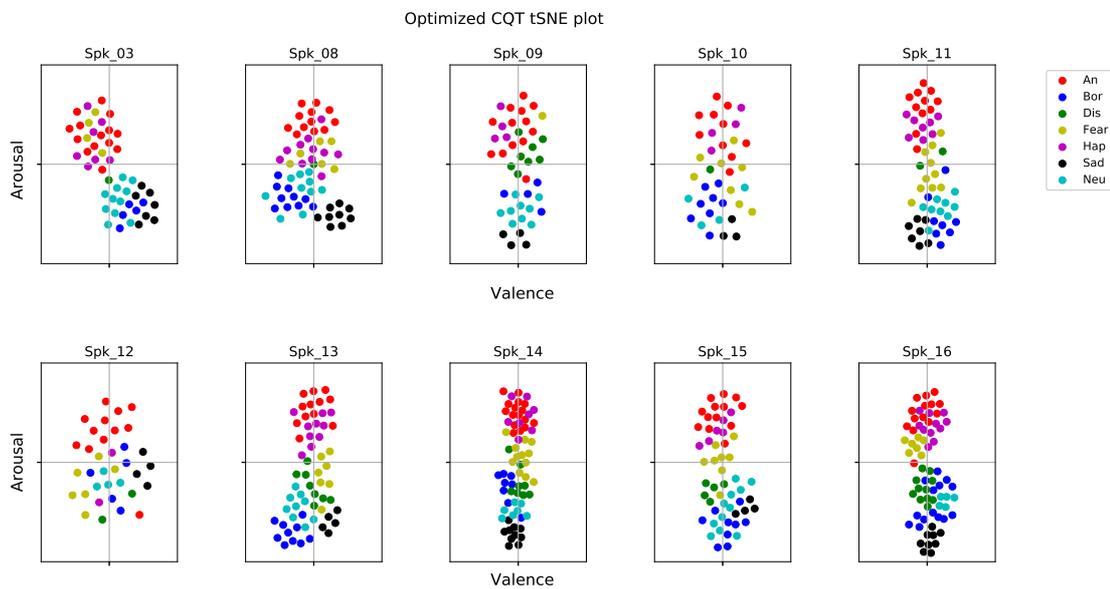}}}
    \end{minipage}%
    \vspace{0.5cm}
    \begin{minipage}{\textwidth}
    \subcaptionbox{t-SNE plot of MFSC embeddings}
    {\hbox{\hspace{-1.5cm}\includegraphics[scale=0.425]{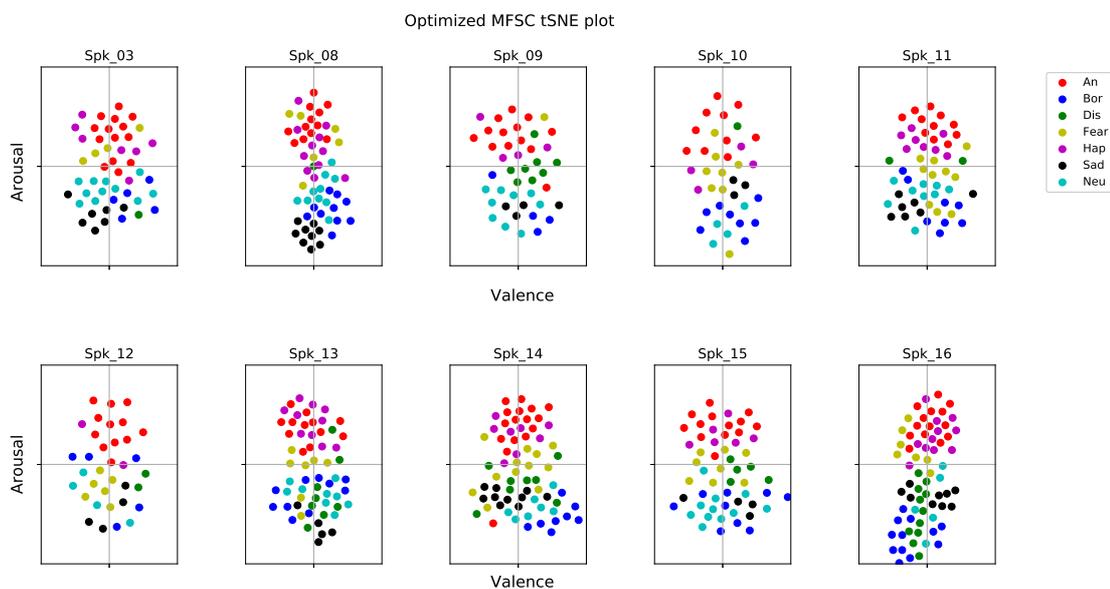}}}
    \end{minipage}%
    \caption{t-SNE plots of CQT and MFSC embeddings for different speakers of EmoDB. The embeddings were extracted from the TDNN architecture. The points in every scatter plot are rotated to keep the \textit{Anger} embeddings on top. The relative positioning of embeddings of other emotions provides an estimate of embedding's spread across the Arousal-Valence axis. Although no particular pattern is observed across the valence (horizontal) axis, embeddings form clusters across the arousal (vertical) scale. CQT embeddings show better clustering of emotions than MFSC.} 
    \label{fig:tSNE_plots}
\end{figure} 

For SER evaluation, we employed commonly used accuracy and UAR as performance metrics. Accuracy is calculated by finding the ratio between the number of correctly classified utterances to the total number of utterances in test set. The UAR metric is given as~\cite{rosenberg2012classifying}:

\begin{equation}
    \mathrm{UAR} = \frac{1}{K} \sum_{i=1}^{K} \frac{A_{ii}}{\sum_{j=1}^{K} A_{ij}}
\end{equation}
 
\noindent where, $A$ refers to the contingency matrix, $A_{ij}$ corresponds to the number of samples in class $i$ classified into class $j$, and $K$ is the total number of classes. As accuracy is considered \emph{unintuitive} for databases with unequal samples across different classes, we optimized the feature extraction parameters based on the UAR metric.

\section{Results and discussion}
\label{results}
\subsection{Embedding visualization}
To understand the discriminability among the constant-Q and mel-scaled representations, we present the tSNE plots of the speech embeddings of CQT and MFSC extracted from the statistics pooling layer of the TDNN architecture with EmoDB database in Fig.~\ref{fig:tSNE_plots}. We kept the \emph{Angry} embeddings on top and considered the positions of other emotion embeddings across the x- and y-axis as separation across valence and arousal axes on the arousal-valence plane~\cite{el2011survey}. Notice better clustering by CQT embeddings at the emotion level. Nevertheless, both the embeddings show discrimination mainly across the arousal (vertical) axis. No major separation across the valence (horizontal) scale is observed across the embeddings. This indicates that the time-frequency representation can distinguish emotions across the arousal scale more than the valence scale. 
\begin{table*}[t!]
\centering
\renewcommand{\arraystretch}{1.5}
\caption{SER performance (in \%) of CQT, MFSC, and CWT-based systems with different classifier back-ends. The \textbf{boldface} values show the highest performance metric obtained over different databases. Figure~\ref{perf_com_res} provides a visual representation of the values in this table.}
\begin{adjustbox}{width=0.8\columnwidth, center}
\begin{tabular}{|c|c|cccccc|}
\hline

\multirow{2}{*}{\textbf{Database}}  & \multirow{2}{*}{\textbf{Architecture}} & \multicolumn{3}{c||}{\textbf{Accuracy}} & \multicolumn{3}{c|}{\textbf{UAR}}\\  [0.25cm]
                           &                               & \multicolumn{1}{c}{\textbf{CQT}} & \multicolumn{1}{c}{\textbf{MFSC}}  & \multicolumn{1}{c||}{\textbf{CWT}} & \multicolumn{1}{c}{\textbf{CQT}}  & \multicolumn{1}{c}{\textbf{MFSC}} & \textbf{CWT} \\ \hline \hline
                           
\multirow{7}{*}{EmoDB}     & Conv2D                        & \multicolumn{1}{c|}{67.15} &  \multicolumn{1}{c|}{52.97} & \multicolumn{1}{c||}{66.60} & \multicolumn{1}{c|}{58.75} & \multicolumn{1}{c|}{48.74} & \textbf{62.24}\\ 
                           & TDNN                         & \multicolumn{1}{c|}{64.20} & \multicolumn{1}{c|}{51.27} & \multicolumn{1}{c||}{63.31} & \multicolumn{1}{c|}{55.97} & \multicolumn{1}{c|}{49.73} & 57.16\\ 
                           & \begin{tabular}[c]{@{}c@{}}Conv2D~+~LSTM~+~Attention \\ (MERC 2020)\end{tabular}            & \multicolumn{1}{c|}{54.25} & \multicolumn{1}{c|}{45.90} & \multicolumn{1}{c||}{55.83} & \multicolumn{1}{c|}{49.96} & \multicolumn{1}{c|}{42.62} &  50.01 \\ 
                           & \begin{tabular}[c]{@{}c@{}}Attention-based LSTM \end{tabular}            & \multicolumn{1}{c|}{61.44} & \multicolumn{1}{c|}{42.83} & \multicolumn{1}{c||}{59.08} & \multicolumn{1}{c|}{57.57} & \multicolumn{1}{c|}{40.92} & 53.97 \\
                           & \begin{tabular}[c]{@{}c@{}}Transformer encoder model \end{tabular}            & \multicolumn{1}{c|}{48.89} & \multicolumn{1}{c|}{46.16} & \multicolumn{1}{c||}{51.21} & \multicolumn{1}{c|}{42.32} & \multicolumn{1}{c|}{41.39} &  45.52 \\ 
                           & \begin{tabular}[c]{@{}c@{}}ECAPA-TDNN \end{tabular}            & \multicolumn{1}{c|}{54.71} &  \multicolumn{1}{c|}{48.30} & \multicolumn{1}{c||}{52.93} & \multicolumn{1}{c|}{47.21} & \multicolumn{1}{c|}{44.98} &  45.89 \\ 
                           & Conv2D-LSTM         & \multicolumn{1}{c|}{65.69} & \multicolumn{1}{c|}{55.51} & \multicolumn{1}{c||}{\textbf{67.36}} & \multicolumn{1}{c|}{60.45} & \multicolumn{1}{c|}{50.08} & 61.76 \\ \hline \hline
\multirow{7}{*}{RAVDESS}   & Conv2D                        & \multicolumn{1}{c|}{39.16}     & \multicolumn{1}{c|}{38.68}     &   \multicolumn{1}{c||}{38.05}   & \multicolumn{1}{c|}{36.96}  &  \multicolumn{1}{c|}{35.70}   &   \multicolumn{1}{c|}{35.64}   \\ 
                           & TDNN                          & \multicolumn{1}{c|}{40.34}     & \multicolumn{1}{c|}{35.00}     &   \multicolumn{1}{c||}{35.00}   & \multicolumn{1}{c|}{36.74}    &  \multicolumn{1}{c|}{30.89}  & 31.23  \\ 
                           & \begin{tabular}[c]{@{}c@{}}Conv2D~+~LSTM~+~Attention  \\ (MERC 2020)\end{tabular}            & \multicolumn{1}{c|}{37.51}     & \multicolumn{1}{c|}{32.40}    &   \multicolumn{1}{c||}{37.56}   & \multicolumn{1}{c|}{35.24}    &   \multicolumn{1}{c|}{31.60}  & 35.21  \\ 
                           & \begin{tabular}[c]{@{}c@{}}Attention-based LSTM \end{tabular}            & \multicolumn{1}{c|}{43.12} & \multicolumn{1}{c|}{32.08} &  \multicolumn{1}{c||}{42.50}  & \multicolumn{1}{c|}{42.74} & \multicolumn{1}{c|}{31.62} &  40.58\\ 
                           & \begin{tabular}[c]{@{}c@{}}Transformer encoder model \end{tabular}            & \multicolumn{1}{c|}{35.69} & \multicolumn{1}{c|}{29.93} &  \multicolumn{1}{c||}{34.16} & \multicolumn{1}{c|}{33.23} &  \multicolumn{1}{c|}{27.96} &  32.28\\ 
                           & \begin{tabular}[c]{@{}c@{}}ECAPA-TDNN \end{tabular}            & \multicolumn{1}{c|}{32.43} & \multicolumn{1}{c|}{34.37} &  \multicolumn{1}{c||}{34.86} & \multicolumn{1}{c|}{31.82} & \multicolumn{1}{c|}{33.61} &  34.16\\ 
                           & Conv2D-LSTM            & \multicolumn{1}{c|}{\textbf{46.94}} & \multicolumn{1}{c|}{42.56} & \multicolumn{1}{c||}{46.43} & \multicolumn{1}{c|}{44.15} &  \multicolumn{1}{c|}{39.80} &  \textbf{44.60} \\ \hline \hline
\multirow{7}{*}{eNTERFACE} & Conv2D                        & \multicolumn{1}{c|}{46.77}     &  \multicolumn{1}{c|}{41.86}    &   \multicolumn{1}{c||}{49.09}   &   \multicolumn{1}{c|}{41.31}   &   \multicolumn{1}{c|}{37.77}   &   43.35 \\ 
                           & TDNN                          & \multicolumn{1}{c|}{52.65}     &  \multicolumn{1}{c|}{35.72}    &  \multicolumn{1}{c||}{56.91}    &   \multicolumn{1}{c|}{39.27}   &   \multicolumn{1}{c|}{33.05}   &  41.92  \\ 
                           & \begin{tabular}[c]{@{}c@{}}Conv2D~+~LSTM~+~Attention \\ (MERC 2020)\end{tabular}            & \multicolumn{1}{c|}{45.33}     &  \multicolumn{1}{c|}{44.58}    &  \multicolumn{1}{c||}{45.49}    &   \multicolumn{1}{c|}{43.35}  &   \multicolumn{1}{c|}{43.07}  &  43.82  \\  
                           & \begin{tabular}[c]{@{}c@{}}Attention-based LSTM \end{tabular}            & \multicolumn{1}{c|}{41.10} &  \multicolumn{1}{c|}{40.02} & \multicolumn{1}{c||}{42.92}  & \multicolumn{1}{c|}{40.43}  &  \multicolumn{1}{c|}{39.16}  & 42.61 \\
                           & \begin{tabular}[c]{@{}c@{}}Transformer encoder model \end{tabular}            & \multicolumn{1}{c|}{38.87} &  \multicolumn{1}{c|}{41.55}  & \multicolumn{1}{c||}{44.78} & \multicolumn{1}{c|}{37.96} & \multicolumn{1}{c|}{38.69} &  42.01 \\
                           & \begin{tabular}[c]{@{}c@{}}ECAPA-TDNN \end{tabular}            & \multicolumn{1}{c|}{55.85} & \multicolumn{1}{c|}{52.73}  & \multicolumn{1}{c||}{54.73} & \multicolumn{1}{c|}{\textbf{55.10}} & \multicolumn{1}{c|}{52.08} & 54.50 \\ 
                           & Conv2D-LSTM            & \multicolumn{1}{c|}{48.36} & \multicolumn{1}{c|}{47.39}  & \multicolumn{1}{c||}{\textbf{59.05}} & \multicolumn{1}{c|}{46.76} & \multicolumn{1}{c|}{44.84} & 54.80 \\ \hline \hline
\multirow{7}{*}{IEMOCAP}   & Conv2D                        & \multicolumn{1}{c|}{53.08}    &  \multicolumn{1}{c|}{48.77}   &   \multicolumn{1}{c||}{55.38}    &  \multicolumn{1}{c|}{45.04}  &  \multicolumn{1}{c|}{40.20}  &  45.85  \\ 
                           & TDNN                          & \multicolumn{1}{c|}{53.45}     &  \multicolumn{1}{c|}{52.36}    &   \multicolumn{1}{c||}{55.25}   &  \multicolumn{1}{c|}{42.07}   &   \multicolumn{1}{c|}{40.14}  &  42.40  \\ 
                           & \begin{tabular}[c]{@{}c@{}}Conv2D~+~LSTM~+~Attention \\ (MERC 2020)\end{tabular}            & \multicolumn{1}{c|}{50.16}     &  \multicolumn{1}{c|}{47.16}   &  \multicolumn{1}{c||}{50.67}  &   \multicolumn{1}{c|}{43.42}   & \multicolumn{1}{c|}{39.78}    &  44.14 \\ 
                           & \begin{tabular}[c]{@{}c@{}}Attention-based LSTM \end{tabular}            & \multicolumn{1}{c|}{47.41} & \multicolumn{1}{c|}{45.68}  & \multicolumn{1}{c||}{51.50} & \multicolumn{1}{c|}{44.22} & \multicolumn{1}{c|}{40.41} &  46.73 \\ 
                           & \begin{tabular}[c]{@{}c@{}}Transformer encoder model \end{tabular}            & \multicolumn{1}{c|}{48.77} & \multicolumn{1}{c|}{43.83} & \multicolumn{1}{c||}{51.45} & \multicolumn{1}{c|}{41.40} & \multicolumn{1}{c|}{38.73} &  42.53 \\ 
                           & \begin{tabular}[c]{@{}c@{}}ECAPA-TDNN \end{tabular}            & \multicolumn{1}{c|}{50.32} & \multicolumn{1}{c|}{46.19}  & \multicolumn{1}{c||}{48.33} & \multicolumn{1}{c|}{42.77} & \multicolumn{1}{c|}{40.90} & 42.53 \\ 
                           & Conv2D-LSTM            & \multicolumn{1}{c|}{54.83} & \multicolumn{1}{c|}{48.34} & \multicolumn{1}{c||}{\textbf{57.31}} & \multicolumn{1}{c|}{46.53} & \multicolumn{1}{c|}{41.11} & \textbf{47.94}  \\ \hline
\end{tabular}
\end{adjustbox}
\label{res_table}
\end{table*}

\begin{figure*}[t]
    \centering
    \hbox{\hspace{-1.65cm}\includegraphics[scale=0.45]{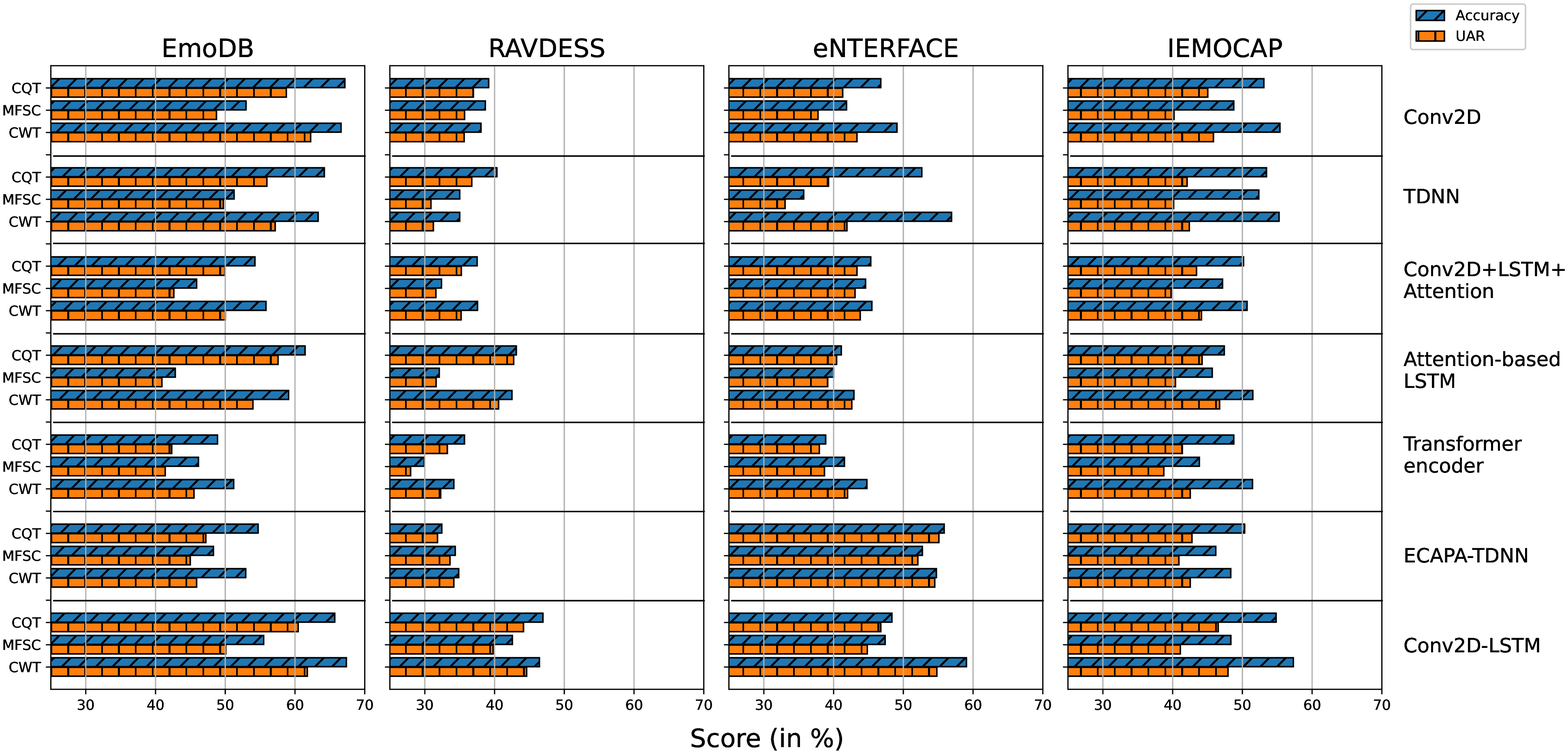}}
    \caption{Bar graph showing accuracy and UAR obtained with different features, neural network architectures, and databases. This is the visual representation of Table~\ref{res_table}.}
    \label{perf_com_res}
\end{figure*}

\subsection{Comparison of SER performances}
\label{sec:comp_ser}
Table~\ref{res_table} shows the results obtained for different features, while Fig.~\ref{perf_com_res} provides the visual representation of the numerical values in the table. Both constant-Q filterbank representations (CQT and CWT) perform better than MFSC in predicting emotion classes across different databases. This demonstrates the inappropriateness of mel-scale for emotion prediction. The considerable improvement in constant-Q representations across different architectures justifies the salience of low-frequency regions of speech for SER. Between the constant-Q features, the performance varies across databases. CWT is either equivalent to or better than CQT over different databases. For EmoDB, CWT mostly shows higher UAR than CQT. For RAVDESS, CQT is better than CWT, whereas, in IEMOCAP and eNTERFACE, CWT performs better than CQT. Over RAVDESS database, an anomaly is the performance of CWT which is equivalent to MFSC with TDNN and Conv2D architectures.

Among the TDNN, Conv2D, and Conv2D-LSTM architectures, Conv2D-LSTM fairs better over every database. Between TDNN and Conv2D, the latter performs better in terms of UAR except in EmoDB with the MFSC feature. This shows a better capability of convolution layers in improving prediction accuracy across all emotions instead of focusing on the more dominant emotion classes. The 2-dimensional convolution in Conv2D utilizes the correlation of features across both time and frequency domains and extracts variations across smaller time-frequency windows defined by the filter size at every layer. This helps accumulate refined emotion information from filters of varying sizes across different layers of the model, which generates a better representation of emotion at the output. The LSTM layer further extracts the required temporal information from the activation of Conv2D layers, thereby improving performance. 

For every architecture, constant-Q features outperform MFSC over every database. Also, the selected state-of-the-art deep learning architectures, i.e., Attention-based LSTM, Transformer encoder model, and ECAPA-TDNN, show inferior performance compared to plain Conv2D, TDNN, and Conv2D-LSTM. The Transformer encoder model has the poorest results among state-of-the-art networks across different databases because of the presence of only non-linear transformation-based attention layers and the lack of filter-based convolution operation to extract both time and frequency-based features. However, Attention-based LSTM architecture provides comparable performance (at least in UAR) to plain Conv2D and TDNN architectures. The final time and frequency-based attention layers in Attention-based LSTM help focus more on the emotion-relevant time and frequency cues. Similarly, ECAPA-TDNN outperforms both plain Conv2D and TDNN on eNTERFACE (in UAR) and provides comparable results on IEMOCAP. As ECAPA-TDNN contains many network parameters, its higher performance on databases with more data samples is justified. The Conv2D-LSTM model combines convolutional and temporal information extraction (due to the recurrent layer) to provide better performance on most databases.

\begin{figure}
    \centering
    \includegraphics[scale=0.4]{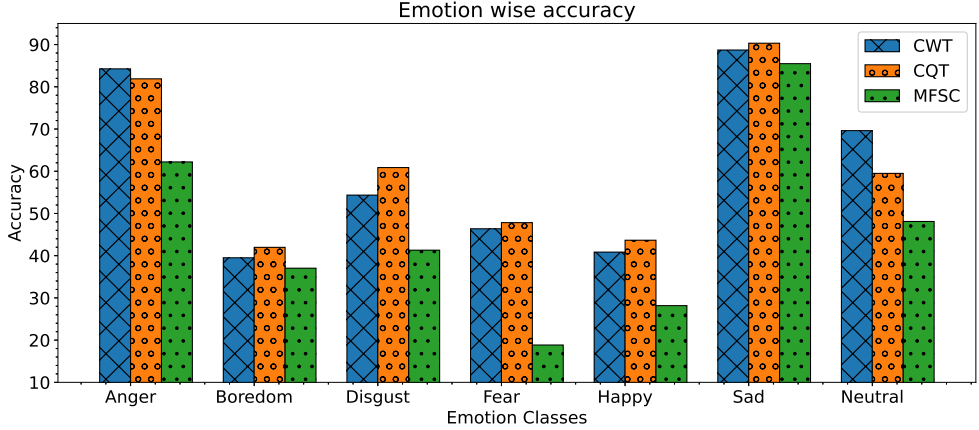}
    \caption{Emotion-wise performance comparison among CQT, CWT, and MFSC on EmoDB database. The accuracies are computed with Conv2D architecture as classifier. Constant-Q features provide greatest advantage over MFSC on high arousal emotion (e.g., \emph{Fear, Anger}).}
    \label{emo-wise-comp}
\end{figure}

Figure~\ref{emo-wise-comp} shows the comparison of emotion-wise accuracy obtained with different features over plain Conv2D architecture. The performance of constant-Q features is similar across different emotions, except for \emph{Neutral} and \emph{Disgust}. As the CWT includes framing and windowing, the emotions less sensitive to high-frequency content and benefited from high-frequency averaging are emphasized by CWT. Figure~\ref{emo-wise-comp} shows that \emph{Anger}, which is known to have greater high frequency relevance~\cite{banse1996acoustic, williams1972emotions}, gains slightly from time-invariance (or averaging) applied at high frequency. The same is observed for \emph{Neutral}. In contrast, \emph{Disgust}, which is also sensitive to high-frequency variations (\cite{banse1996acoustic}), is represented better with CQT rather than with averaged high-frequency representation in CWT.

The improvement with constant-Q features is more than that for MFSC, especially on high-arousal emotions (\emph{Fear, Anger, Happy}) because of the higher pitch resolution in the former. Contours of pitch harmonics in high-arousal emotions contain sudden rises and drops. These intonations are better captured in constant-Q features with higher resolution at low frequencies. For low-arousal emotions like \emph{Sad} and \emph{Boredom}, pitch contours usually follow straight-line patterns causing similar performance of MFSC and constant-Q features. For every feature, \emph{Sad} emotion class yields the highest classification accuracy. MFSC provided the lowest accuracy score for \emph{Fear} whereas the constant-Q representations yield the lowest accuracy score for \emph{Boredom}. The difference between recognition accuracies of \emph{Sad} and \emph{Boredom} appears mainly because of similar arousal and valence characteristics.

\begin{table}[t!]
    \centering
    \renewcommand{\arraystretch}{1.2}
    \caption{Performance comparison with relevant works. Abbreviations: eGeMAPS~=~extended Geneva minimalistic acoustic parameter set, SVM~=~Support vector machine, COVAREP~=~Collaborative voice analysis repository, ECAPA-TDNN~=~Emphasized channel attention, propagation and aggregation in TDNN, MFLE~=~Mel-frequency log energy, CLDNN~=~Convolutional bidirectional LSTM DNN, LOSGO~=~Leave one speaker group out.}
    \begin{adjustbox}{width=\textwidth, center}
    \begin{tabular}{cccc}
    \hline

        \textbf{Reference} & \textbf{Methodology}   & \textbf{Evaluation method}   & \begin{tabular}[c]{@{}c@{}} \textbf{Acc./UAR} \\ (in \%) \end{tabular}  \\ \hline

        \multicolumn{4}{l}{\emph{EmoDB}} \\ \hline \hline
        \citeauthor{parry2019analysis}~(\citeyear{parry2019analysis})   & \begin{tabular}[c]{@{}c@{}} MFSC feature, \\ CNN-LSTM classifier  \end{tabular}  & \begin{tabular}[c]{@{}c@{}} 80:10:10\% train/valid/test split, \\ 7 emotion classes  \end{tabular}  &  -/69.72  \\[0.5cm] 
        
        \citeauthor{Triantafyllopoulos2019}~(\citeyear{Triantafyllopoulos2019}) & \begin{tabular}[c]{@{}c@{}} eGeMAPS, ComParE features, \\ SVM classifier
        \end{tabular}   & \begin{tabular}[c]{@{}c@{}}LOSO cross-validation, \\ 7 emotion classes \end{tabular} &   \begin{tabular}[c]{@{}c@{}} eGeMAPS:~-/61.04 \\ ComParE:~-/72.34 \end{tabular} \\ [0.5cm]
        
        \citeauthor{HAIDER2021101119}~(\citeyear{HAIDER2021101119})  & \begin{tabular}[c]{@{}c@{}} eGeMAPS features, \\ No feature selection, \\ SVM classifier \end{tabular} &  \begin{tabular}[c]{@{}c@{}} LOSO cross-validation, \\ 7 emotion classes \end{tabular}  &  -/68.5 \\[0.7cm]
        
         \begin{tabular}[c]{@{}c@{}} \textbf{Ours} \\ (Best result) \end{tabular} &   \begin{tabular}[c]{@{}c@{}} CWT feature, \\ Conv2D-LSTM classifier \end{tabular}  & \begin{tabular}[c]{@{}c@{}} LOSO cross-validation, \\ 7 emotion classes  \end{tabular} &   67.36/61.76  \\ [0.5cm]
        \hline

        \multicolumn{4}{l}{\emph{RAVDESS}} \\ \hline \hline
        \citeauthor{guizzo2020}~(\citeyear{guizzo2020}) & \begin{tabular}[c]{@{}c@{}} Spectrogram applied to multi-time-scale \\ learning with CNN classifier.   \end{tabular} & \begin{tabular}[c]{@{}c@{}} 4-fold cross-validation, 70:20:10\% \\ train:valid:test split, 8 emotion classes \end{tabular} & 55.85/- \begin{tabular}[c]{@{}l@{}}  \end{tabular}\\[0.5cm]
        
        \citeauthor{Dissanayake2020}~(\citeyear{Dissanayake2020})    & \begin{tabular}[c]{@{}c@{}} MFCC feature, \\ CNN with autoencoder \\ and LSTM classifier \end{tabular}   & \begin{tabular}[c]{@{}c@{}} Training:Valid:Test; 22:1:1 speakers, \\ 8 classes mapped to 3 classes, \\ Positive, Negative, and Neutral \end{tabular} &  -/56.71\\[0.75cm]
        
        \citeauthor{beard-etal-2018-multi}~(\citeyear{beard-etal-2018-multi})    &   \begin{tabular}[c]{@{}c@{}}  COVAREP features, \\ LSTM classifier  \end{tabular} & \begin{tabular}[c]{@{}c@{}} Evaluation method not defined,  \\ 8 emotion classes \end{tabular} &   41.25/-\\[0.6cm]
        
       \begin{tabular}[c]{@{}c@{}} \textbf{Ours} \\ (Best result) \end{tabular}   &   \begin{tabular}[c]{@{}c@{}} CWT feature, \\ Conv2D-LSTM classifier \end{tabular} &  \begin{tabular}[c]{@{}c@{}} LOSO cross-validation, \\ 8 emotion classes  \end{tabular}  &  46.43/44.60 \\ [0.6cm] 
        \hline

        \multicolumn{4}{l}{\emph{eNTERFACE}} \\ \hline \hline         
        \citeauthor{Triantafyllopoulos2019}~(\citeyear{Triantafyllopoulos2019}) & \begin{tabular}[c]{@{}c@{}} eGeMAPS, ComParE features, \\ SVM classifier
        \end{tabular}   & \begin{tabular}[c]{@{}c@{}}LOSO cross-validation, \\ 6 emotion classes \end{tabular} &   \begin{tabular}[c]{@{}c@{}} eGeMAPS:-/47.87 \\ ComParE:-/65.60 \end{tabular} \\ [0.5cm]
        
        \citeauthor{zhang2017speech}~(\citeyear{zhang2017speech})  & \begin{tabular}[c]{@{}c@{}}MFSC feature (no delta, double-delta), \\ DCNN (AlexNet) with average pooling, \\ SVM classifier  \end{tabular} & \begin{tabular}[c]{@{}c@{}} LOSGO cross-validation, \\ 6 emotion classes \end{tabular} &  51.33/- \\[0.7cm]
        
        \begin{tabular}[c]{@{}c@{}} \textbf{Ours} \\ (Best result) \end{tabular}   &  \begin{tabular}[c]{@{}c@{}} CWT feature, \\ Conv2D-LSTM classifier \end{tabular}  & \begin{tabular}[c]{@{}c@{}} LOSO cross-validation, \\ 6 emotion classes  \end{tabular}  &   59.05/54.80  \\ [0.5cm]  \hline

        \multicolumn{4}{l}{\emph{IEMOCAP}} \\ \hline \hline
        \citeauthor{pandey2022attention}~(\citeyear{pandey2022attention})    & \begin{tabular}[c]{@{}c@{}} MFSC feature, \\ CNN-LSTM as classifier \end{tabular} & \begin{tabular}[c]{@{}c@{}} LOSO cross-validation, \\ 4 emotion classes \end{tabular}  &   51.82/-  \\[0.5cm]
        
        \citeauthor{kumawat21_interspeech}~(\citeyear{kumawat21_interspeech})    & \begin{tabular}[c]{@{}c@{}}  MFCC feature, \\ ECAPA-TDNN classifier \end{tabular}  & \begin{tabular}[c]{@{}c@{}} Leave-one-session-out \\ cross validation, \\ 4 emotion classes \end{tabular}  &    58.76/-   \\[0.75cm]
        
        \citeauthor{Dissanayake2020}~\citeyear{Dissanayake2020}    & \begin{tabular}[c]{@{}c@{}} MFCC feature, \\ CNN with autoencoder \\ and LSTM classifier \end{tabular}   & \begin{tabular}[c]{@{}c@{}} 5 sessions in training, half of 6th \\ session in valid, half in test, \\ 4 emotion classes \end{tabular} &  -/46.79  \\[0.75cm]
        
        \citeauthor{Meyer2018}~(\citeyear{Meyer2018})    & \begin{tabular}[c]{@{}c@{}} MFLE feature, \\ CLDNN classifier \end{tabular}   & \begin{tabular}[c]{@{}c@{}} LOSGO cross-validation, \\ 4 emotion classes \end{tabular} &   -/59.5  \\[0.5cm]
        
        \begin{tabular}[c]{@{}c@{}} \textbf{Ours} \\ (Best result) \end{tabular}  &  \begin{tabular}[c]{@{}c@{}} CWT feature, \\ Conv2D-LSTM classifier \end{tabular}  &  \begin{tabular}[c]{@{}c@{}} LOSO cross-validation, \\ 4 emotion classes  \end{tabular}  &   57.31/47.94 \\ [0.5cm]
        \hline

    \end{tabular}
    \end{adjustbox}

    \label{perf_comp_table}
\end{table}

Table~\ref{perf_comp_table} lists a few recent and relevant SER works. The absence of consensus on a standard experimental setup in SER literature is evident in the table. High SER performance can be achieved with less stringent evaluation frameworks, e.g., speaker-dependent train/test split, fewer cross-validation folds, fewer emotion classes from the database, etc. Thus, the performance comparison of multiple systems is difficult and mostly non-conclusive. Other factors like the absence of standard performance metric and lack of reproducible research in SER adds to inaccuracy in comparison. In our attempt to compare the employed system with other SER works, we include both methodology and experimental setup in the table so that the differences among works are understood and, to some extent, the comparison is valid. 

Comparison with the referred works shows that our method outperforms most of them on eNTERFACE and IEMOCAP databases but not EmoDB and RAVDESS. In EmoDB, the less stringent evaluation protocol in~\cite{parry2019analysis} and~\cite{HAIDER2021101119}, in terms of train/test split, provides better classification results. The same argument applies to the works on the RAVDESS database. Although higher results can be achieved with lenient evaluation strategies, they do not imitate the real-world SER testing scenario. As discussed in Section~\ref{train_test_eval}, the LOSO cross-validation extensively evaluates the generalizability of a system with input from different unseen speakers without a large increase in computation complexity with small databases, making it a better evaluation strategy, especially for SER. 

Let us now consider the works that use large feature sets~\cite{Triantafyllopoulos2019, HAIDER2021101119}. The famous eGeMAPS~\cite{gemaps} and ComParE~\cite{schuller2016interspeech} feature sets contain many handcrafted features, including a combination of spectral and prosody features with statistics of a different order, unlike our method, which used spectral information only. Although considered appropriate as baseline, eGeMAPS and ComParE feature sets were designed after reviewing and selecting various handcrafted features found successful in previous SER studies~\cite{gemaps}. Therefore, using these feature sets can also be considered human intervention in the train/test phases of the machine learning system contrary to using a specific handcrafted time-frequency representation for supervised learning and hence, also explains the performance difference from our method.

\begin{figure}[t!]
    \centering
    \begin{subfigure}{0.556\textwidth}
    \hbox{\hspace{-1cm}\includegraphics[scale=0.35]{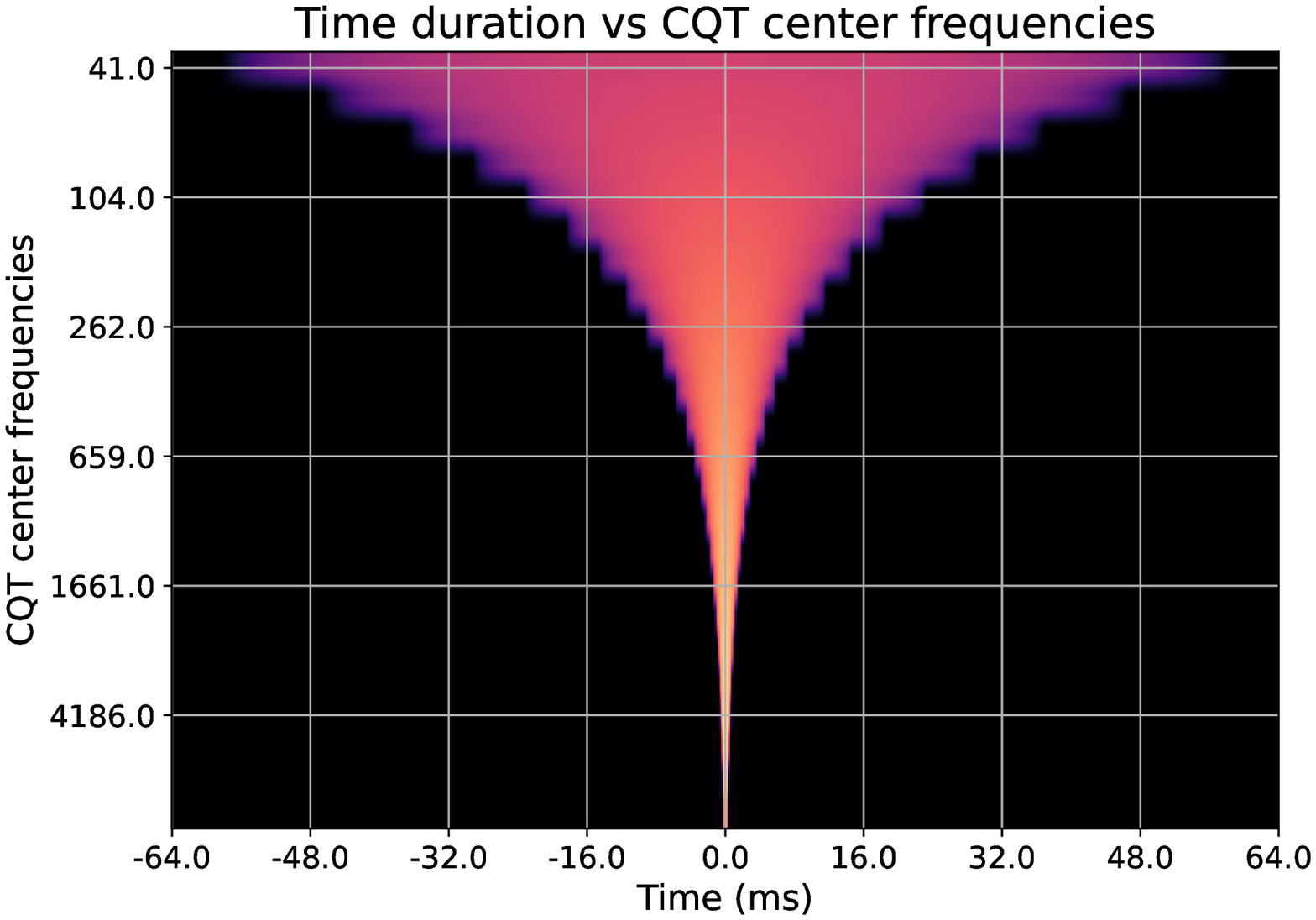}}
    \end{subfigure}%
    \begin{subfigure}{0.55\textwidth}
    \hbox{\hspace{1cm}\includegraphics[scale=0.35]{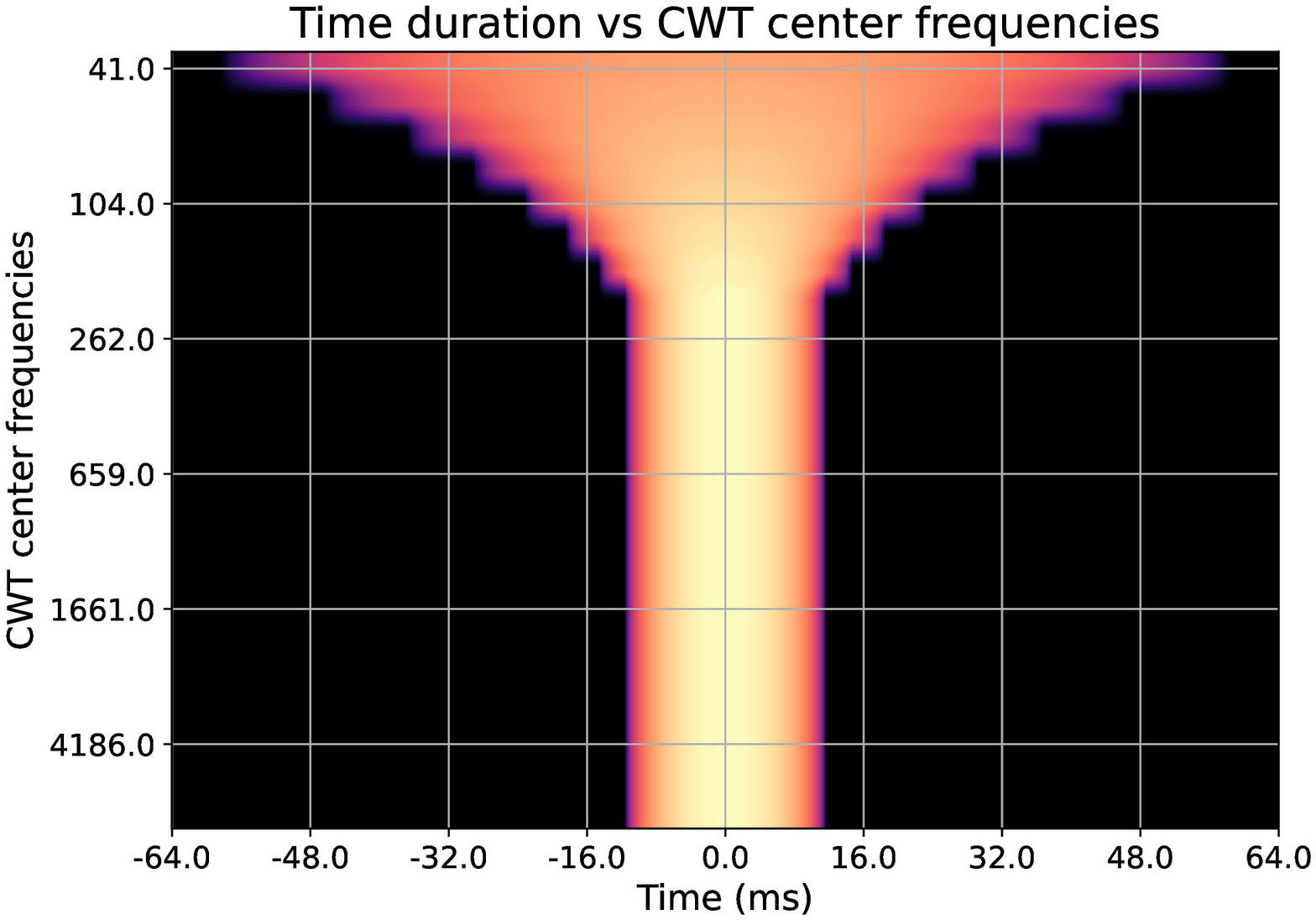}}
    \end{subfigure}
    \begin{subfigure}{0.5\textwidth}
    \includegraphics[scale=0.354]{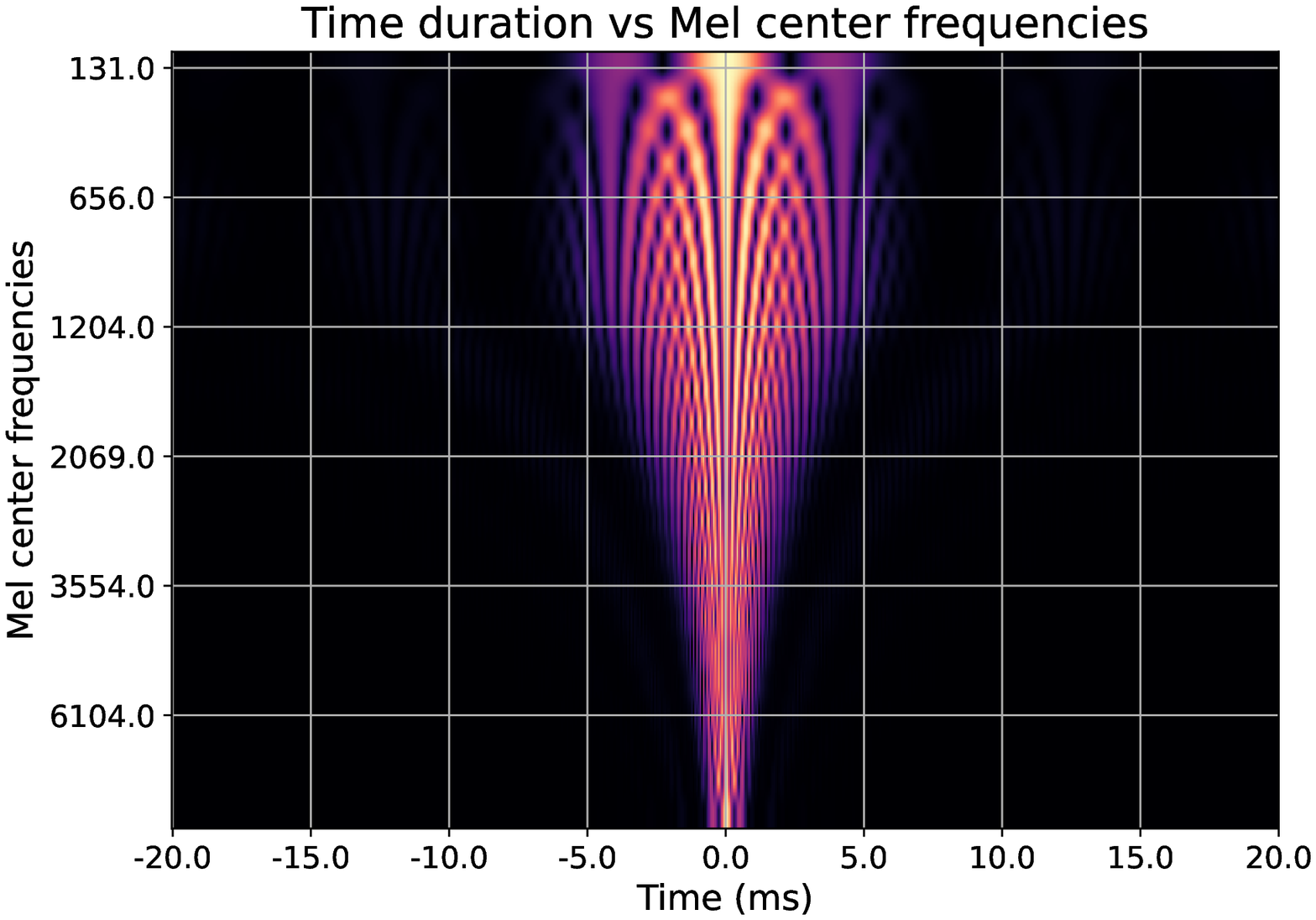}
    \end{subfigure}
    \caption{Time invariance provided by CQT, CWT, and MFSC with changing frequency. y-axis: the center frequency of filters (or frequency bins), $24$ filters per feature; x-axis: the filters' corresponding time spread. Both CQT and CWT use basis functions with similar time spread. However, framing in CWT makes time invariance fixed to $20$~ms at mid and high-frequency bins. Mel-filters have a much lower time spread than constant-Q filters at low frequencies.}
    \label{fig:time_invar}
\end{figure}

\subsection{Time-invariance analysis over features}
The increased low-frequency resolution in constant-Q filterbank representation brings out better resolution of pitch information (Section~\ref{freq_domain_comp}: Fig.~\ref{pitch_harmonic_sep}) thereby enhancing emotion recognition performance (Section~\ref{sec:comp_ser}: Table~\ref{perf_comp_table}, Fig.~\ref{perf_com_res}, and Fig.~\ref{emo-wise-comp}). Regarding the difference in time-invariant property between MFSC and constant-Q-based responses, we compared the temporal lengths of optimized CQT basis and mel-filters in Fig.~\ref{fig:time_invar}. The low-frequency CQT basis functions (bases with center frequency $\leq 262$~Hz) have a temporal spread greater than that of the $20$~ms window used in STFT. Also, the modulus operation on computed CQT coefficients removes the phase information, leading to time-invariance defined by the length of the basis functions. Human pitch frequencies exist at around $250$~Hz~\cite{hillenbrand2009role}. Therefore, the time-invariance in CQT is greater than that of the standard STFT around pitch frequencies, making the former more robust against emotion-irrelevant pitch variations. Notice that the temporal spread of mel-filters remains below $20$~ms over the complete frequency range. Hence, in MFSC, the temporal invariance is defined by the window length ($\phi$) fixed to $20$~ms. In CWT, we applied a similar averaging of computed CWT coefficients by a $20$~ms window causing a varying time-invariance when the CWT basis is greater than the window length and fixed time-invariance otherwise. The results show that the combination of varying and fixed time-invariance leads to slight improvement across most emotion-database pairs. Additionally, increased time-resolution at high frequencies in constant-Q filterbank response contributes less to SER. Rather, capturing long temporal information is a better approach. The long time-scale information alongside the translation invariant feature learning capability of deep networks (obtained from convolutional layers, statistics pooling layer, etc.) improved SER performance in our experiments. 

\begin{table}[t]
\renewcommand{\arraystretch}{1.5}
\captionof{table}{FLOPs and time required for computation of different features. MFLOPs~=~Mega FLOPs.}
\begin{adjustbox}{width=0.65\textwidth, center}
\begin{tabular}{c|ccc}
\hline
\begin{tabular}{@{}c@{}} \textbf{Complexity} \\ \textbf{Parameter} \end{tabular}  & \textbf{MFSC}   & \textbf{CQT}    &  \textbf{CWT}     \\ \hline
\hline
FLOPs   & 8.60~MFLOPs  & 6.73~MFLOPs  & 110.88~MFLOPs  \\ 
Time   & 1.047~s  & 1.629~s & 1.118~s  \\ 
\hline
\end{tabular}
\end{adjustbox}
\label{time_comp}
\end{table}

Although our CWT implementation is very similar to the first layer scattering coefficients~\cite{anden2014} (Section~\ref{comp_tf_rep}), unlike the former, the scattering coefficients provide fixed time-invariance over the complete frequency range. In a separate study, we found that our CWT implementation also performs better than the first layer scattering coefficients for SER~\cite{singh2021deep}. 

\subsection{Complexity analysis}
To compare the complexity of different features, we calculated the floating-point operations (FLOPs) and the time required to compute the features. Table~\ref{time_comp} reports features' FLOPs and computation time averaged across 100 runs of feature extraction of a randomly chosen one-second speech sample. The FLOPs count estimates the number of mathematical operations (additions, subtractions, multiplications, and divisions) required to compute the feature. We calculated FLOPs during feature extraction using only one CPU core (Intel Xeon E5-2670 $2.6$~GHz) and Linux \texttt{perf}\footnote{\url{https://perf.wiki.kernel.org/index.php/Main_Page}} command. The table shows that the number of FLOPs and computation time is inconsistent across features. For example, CQT requires fewer FLOPs but more time to compute, while CWT requires more FLOPs but less computation time. This inconsistency might result from how processors compute integers, fractions, etc., and the time it takes to perform such computations. The CQT requires fewer floating-point operations, but they might take more time to compute, whereas CWT needs more operations but might take less time to calculate.

\section{Conclusion}
\label{sec9}
This paper presented the role of constant-Q filterbank based time-frequency representations, namely CQT and CWT, for SER and compared them with the traditional mel-scaled representation. Our analysis expounded the emotion-relevant advantages provided by the former representations in both time and frequency domains. The comparison with the latter also showed that the greater emphasis provided by the constant-Q non-linearity over low-frequency regions accounts for better suitability of constant-Q representations for SER. The superiority remained consistent across different neural network architectures. From the experiments and analysis, we conclude the following:

\begin{enumerate}
  
    \item Constant-Q filterbank representation provides higher time-invariance and increased frequency resolution at low speech-frequencies, causing an improved SER performance compared to mel-scale based representation.
    
    \item In the time-domain, CQT provides time-invariance increasing towards low frequencies leading to robustness against emotion irrelevant temporal variations and eventually better emotion prediction.
    
    \item The CWT implementation bears a combination of varying and fixed time-invariance over different frequency bins and offers an advantage similar to CQT in SER performance. The difference in performance between the two is attributed to the fixed time-invariance in CWT at mid and high frequencies.
    
    \item Like mel-scale representation, constant-Q representations also provide stability to time-warp deformations begetting a robust descriptor in terms of variations due to different speaking styles.
    
    \item In the frequency domain, constant-Q filterbank representations are more efficient in resolving pitch harmonics than mel-scaled representations contributing to a better SER performance because of the higher relevance of pitch in emotion prediction.
    
    \item  Constant-Q representations outperform mel-scale representations over multiple neural network architectures.
    
    \item Studies in psychology show better emotion perception abilities of musicians than non-musicians~\cite{lima2011speaking, good2017benefits}. As CQT is a more appropriate representation for music analysis~\cite{brown1991calculation}, better SER capabilities of CQT over MFSC suggest some linkage between human emotion and music perception.
\end{enumerate} 

Although there was a noticeable improvement with constant-Q filterbank representations, there is a need for further effort to develop a deployable real-world SER system. Table~\ref{time_comp} shows approximately $60\%$ and $10\%$ more computation time for CQT and CWT features, respectively, compared with MFSC. This increase is the cost for better performance calling for further exploration of time-frequency representations for SER tasks. Future directions of this work include cross-corpora evaluation to study the generalization ability of constant-Q representation with out-of-domain training data. The SER datasets are small-scale datasets that limit the investigations of large-scale deep architectures for this task. The constant-Q representation can be examined in a transfer learning framework which involves learning a pretext task first on a large dataset followed by training downstream emotion recognition task on the limited size dataset.







\bibliography{mybibfile}

\end{document}